\documentclass[twocolumn, 10pt]{article}
\usepackage{times,amsmath,amssymb,amsthm}
\usepackage{graphicx}
\usepackage{fancyhdr}
\usepackage{graphicx}
\usepackage{geometry}
\usepackage{color}
\usepackage{bbold}
\usepackage{wrapfig}
\usepackage{hyperref}
\usepackage[dvipsnames]{xcolor}
\usepackage{enumerate}
\usepackage{enumitem}
\usepackage{lipsum}
\usepackage[T1]{fontenc}
\usepackage[utf8]{inputenc}
\usepackage[affil-it]{authblk}
\usepackage{blindtext}
\usepackage{abstract}
\usepackage{lipsum}
\usepackage{multicol}

\title{\textbf{{\Large{Emergence of an aperiodic Dirichlet space from the tetrahedral units of an icosahedral internal space}}}\\ 
}
\author[1]{Amrik Sen\thanks{amrik@quantumgravityresearch.org}}
\author[1]{Raymond Aschheim}
\author[1]{Klee Irwin}
\affil[1]{Quantum Gravity Research, Los Angeles, CA 90290, USA.}

\begin{document}
\twocolumn[
  \begin{@twocolumnfalse}
    \maketitle
    \begin{abstract}
      We present the emergence of a root system in six dimensions from the tetrahedra of an icosahedral
core known as the 20-group (20G) within the framework of Clifford's geometric algebra. Consequently, we establish a connection between a three dimensional icosahedral seed, a six dimensional Dirichlet quantized host and a higher dimensional lattice structure. The 20G, owing to its icosahedral symmetry, bears the signature of a 6D lattice that manifests in the Dirichlet integer representation. We present an interpretation whereby the three dimensional 20G can be regarded as the core substratum from which the higher dimensional lattices emerge. This emergent geometry is based on an induction principle supported by the Clifford multivector formalism of 3D Euclidean space. This lays a geometric framework for understanding several physics theories related to $SU(5)$, $E_6$, $E_8$ Lie algebras and their composition with the algebra associated with the even unimodular lattice in $\mathbb{R}^{3,1}$. The construction presented here is inspired by Penrose's \textit{three world} model. \vspace{1cm}
  
    \end{abstract}
  \end{@twocolumnfalse}
  ]
  \saythanks
\section{Introduction}
This paper investigates the emergence of higher dimensional $E_8$ lattice geometry from the spatially physical 3D world of icosahedral space.  This supports the case for 3D being the breeding ground for understanding principles which attempt to unify the laws of nature. This mathematical structure is encoded here within the language of Lie algebra.

We present here the induction of a six-dimensional (6D) embedding of lattice geometry, $\Lambda_{{D^{3\bigoplus \phi3}_6}} := \Lambda_{D_3} \bigoplus \phi \Lambda_{D_3}$, from the icosahedral symmetry group of a 20-tetrahedra (20G) cluster that is at the core of a newly discovered 3D quasi-crystalline structure called the \textit{Quassicrystalline Spin Network} \cite{Fang15}. This 6D composite lattice forms the core of an eight dimensional internal space. This internal space manifests physically in a four dimensional Minkowski spacetime, represented here by the unimodular lattice $II^{3,1}$ and together constitutes an aperiodic Dirichlet space. The subscripts in this notation denote the dimension of the root system and the superscripts denote the base dimensions from which the higher root system is deduced. The presence of the golden mean, $\phi$, is a result of Dirichlet decomposition of 3D space and is elucidated in subsequent sections. Moreover, the notation $\Lambda_{D_3}$ denotes the lattice associated with the Lie algebra $D_3$.    

The purpose of the current paper is to demonstrate an inductive framework for the creation of a higher dimensional lattice from the simple components of 3D quasicrystalline forms. 
This approach is distinct from the one by Dechant \cite{Dechant15} where the birth of $E_8$ is deduced through the non-crystallographic intermediary root systems: $H_3$ and $H_4$, but is inspired by Dechant's use of Clifford spinors as an induction tool for creating higher dimensional forms from 3D icosahedra. 
In this paper, the induction is presented through an intermediary host structure. The precise representation of this intermediary host is $\Lambda_{D_3} \bigoplus \phi \Lambda_{D_3} \bigoplus \Pi^{6,0}$ but will often be represented in a concise manner as $\Lambda_{D_3} \bigoplus \phi \Lambda_{D_3} $, where the representation of the physical dimensions is suppressed and only the six dimensional embedding of the internal space is referred to as the intermediary host.    


\subsection{Organization of the paper}
Following the introductory section, we define spinors and shifters in the language of Clifford's geometric algebra in sec. (2). Dirichlet coordinates are also introduced in sec. (2). The construction of the 20G is explained in detail in sec. (2). In sec. (3), we show the emergence of the Dirichlet quantized host in 6D as a consequence of the spawning of dimensions within the Dirichlet integer representation. In sec. (4), we present a pathway to connect the Dirichlet host with a higher dimensional lattice through a series of transformations of Cartan matrices within the framework of Lie theory. We also comment on the properties of these transformations that reveal the emergent geometry. Finally in sec. (5), we provide concluding remarks and discuss the direction of future research.  

\section{Aperiodic internal space and Clifford motors}
In this section, we demonstrate that an icosahedral core comprising a group of twenty tetrahedra is constructed from the vertices of a tetrahedron hosted by a FCC lattice by using Clifford motors\footnote{Motors are the composition of rotors/spinors and translators/shifters.} ({\textbf cf.} p. 132 in \cite{Kanatani15}) with Dirichlet integer coefficients. This is followed by \textit{dilation} of the inter-tetrahedral gaps based on a metric prescribed by an infinite Fibonacci word. The resulting structure has icosahedral symmetry and a quasicrystalline diffraction pattern with a finite number of vertex-types \cite{Fang15, Fang14}. 
\subsection{Clifford spinors in Dirichlet coordinates}
Simply speaking, spinors are rotors. The action of a spinor on a vector results in a rotation of the vector in Euclidean space. In this section, we formally introduce Clifford spinors as our primary rotor mechanism for the geometric construction discussed subsequently. 
\subsubsection{Spinors of $Cl_3$}
\begin{figure}[h!]
\begin{center}
\includegraphics[scale=0.25]{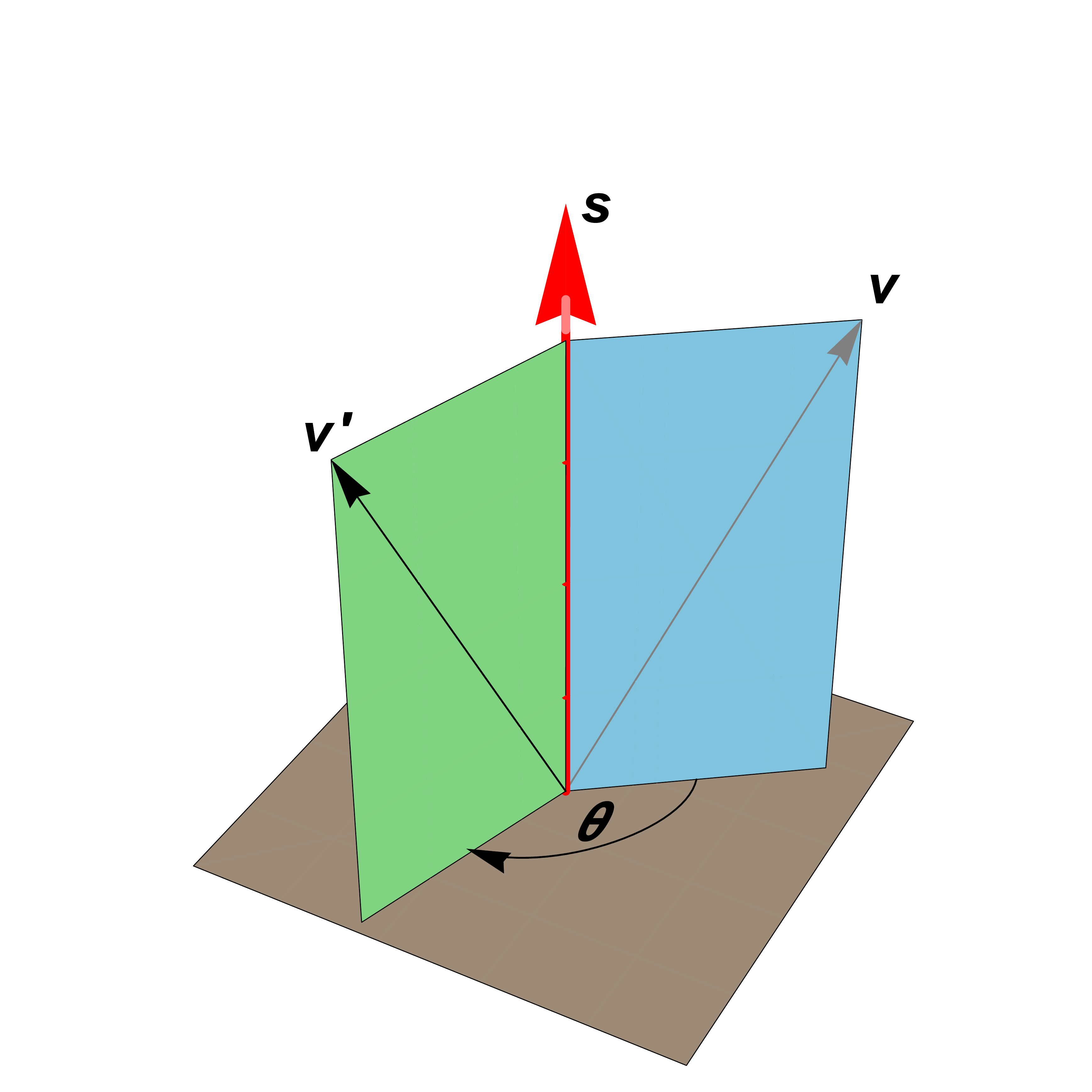}
\end{center}
\caption {The action of a spinor $s$, with axis given by ${\bf s}$, on the vector ${\bf v}$ rotates the latter by an angle $\theta$ resulting in the vector ${\bf v}'$. The length of ${\bf s}$ is proportional to the angle of rotation $\theta$. In this figure, the planes containing vectors (${\bf v}$, ${\bf s}$) and (${\bf v}'$, ${\bf s}$) are orthogonal to the plane of rotation. Moreover, the spinor-axis denoted by the vector ${\bf s}$ is also normal to the plane of rotation.} 
\label{spinor_def}
\end{figure}

The icosahedral core is generated from the simplest \textit{Platonic} solid \cite{Coxeter73} by the action of \textit{spinors}. Spinors are \textit{pure rotors}, i.e., the action of a spinor on a vector results in the rotation of the vector by a desired angle around the axis defined by the spinor. As mentioned, the starting point of our inductive framework is the 3D Euclidean space defined by $\mathbf{R}^3$. In the language of Clifford's geometric algebra, we work within the algebra of $Cl_3$ of $\mathbf{R}^3$. Therefore, we use the 3D spinor, $s$, defined by the elements of the even sub-algebra of $Cl_3^+$ over the subspace $\mathbf{R}  \bigoplus \bigwedge^2 \mathbf{R}^3$ \cite{Lounesto01}. Thus, 
\begin{equation}
s = s_0 + s_1 {\bf e}_{23} + s_2 {\bf e}_{31} + s_3 {\bf e}_{12} \equiv {\bf s}{\bf e}_{123},
\label{spinor_def1}
\end{equation}
where the axis of rotation (of the spinor) is defined by the axis-vector ${\bf s} = \tilde{s}_1 {\bf e}_1 + \tilde{s}_2{\bf e}_2 + \tilde{s}_3{\bf e}_3$. We note that $({\bf s}{\bf e}_{123})^2 = -|{\bf s}|^2$. The magnitude (length) of the bivector (hence that of the spinor) is proportional to the angle by which it rotates the object-vector, ${\bf v}$. This is evident by re-writing the spinor in eq~\eqref{spinor_def1} as follows:
\begin{equation}
s = e^{\frac{1}{2}{\bf s}{\bf e}_{123}} = \cos \frac{\theta}{2} + {\bf e}_{123}\hat{{\bf s}} \sin \frac{\theta}{2},
\label{spinor_def2}
\end{equation}
where $\hat{{\bf s}} = \frac{{\bf s}}{|{\bf s}|}$ is the unit vector that denotes the direction of the spinor-axis and the angle of rotation, $\theta = |{\bf s}|$\footnote{Compare the role of the bivector ${\bf s}{\bf e}_{123}$ of $Cl_3$ in $\mathbf{R}^3$ and $i\theta$ in $\mathbf{C}$ in the exponential form demonstrating a rotation. Here, ${\bf e}_{123}\hat{{\bf s}} = - \hat{{\bf s}}{\bf e}_{123}$.}. Finally, the rotation of the vector, ${\bf v} = v_1 {\bf e}_1 + v_2 {\bf e}_2 + v_3 {\bf e}_3$, is given by 
\begin{equation}
{\bf v} \xrightarrow{s} s {\bf v} s^{-1} = s {\bf v} \frac{\bar{s}}{s\bar{s}} = s {\bf v} \bar{s}, 
\label{spinor_rot}
\end{equation}
because $s \in \mathbf{Spin}(3) \implies s\bar{s} = 1$. Moreover, $s$ is a two fold covering which means that the action of $s$ and $-s$ on ${\bf v}$ by the way of $s {\bf v} \bar{s}$ and $(-s) {\bf v} (-\bar{s})$ is identical. The action of the spinor $s$ on a vector ${\bf v}$ is graphically illustrated in Figure (\ref{spinor_def}). 

\subsubsection{Dirichlet coordinates}
\label{D}
Recall that our objective is to generate an icosahedral core that has a characteristic \textit{five fold rotational symmetry}. This entails a representation of the coordinate frame by quadratic integers in the quadratic field $\mathbf{Q}(\sqrt{D}),~D=5$. $D=5$ defines the quadratic integer ring with \textit{five} fold symmetry.\footnote{Note that the three fold symmetry corresponding to the $\frac{2\pi}{3}$ rotations have their axes of rotation passing through the vertices of a dodecahedron that have coordinates in the Dirichlet frame. Hence the Dirichlet integer decomposition used in this paper follows entirely from pentagonal symmetry and the implications of the three fold symmetry in the coordinate representation is accounted for automatically.} The units of this ring corresponding to the solution of \textit{Pell's equation}\footnote{$x^2 - Dy^2 = 1$ with $D=5$} are of the form $\pm \phi^{n},~n \in \mathbf{Z}$ and $\phi = \frac{1+\sqrt{D}}{2}$ is the \textit{golden ratio} when $D=5$. Thus the quadratic integer ring corresponding to the desired pentagonal rotational symmetry has algebraic integers of the form $\mathbf{Z}[\phi] = a + \phi b,~a,b \in \mathbf{Z}$ which are known as  \textit{Dirichlet} integers. The Dirichlet coordinate frame is defined as a set of coordinates that span the ring of Dirichlet integers \cite{Dirichlet28, Kronecker89, KronFuchs97}.\footnote{This ring has also been studied in the context of other objects with five fold symmetry like the \textit{Penrose tilings} (ref. pp. 60-64 in \cite{Bruijn81}).} This decomposition of the coordinates into two orthogonal golden and non-golden parts lies at the core of the \textit{doubling} of dimensions representative of an emergent geometry, as explained in subsequent sections. We use the notations $a + \phi b$ and $(a,b)$ interchangeably, both implying the presence of two orthogonal coordinate frames, viz. non-golden (first component) and golden (second component) coordinate frames. Therefore, the symmetry group of the icosahedron ensures, \textit{a priori}, that the Dirichlet coordinates are sufficient to host all the vertices of the 20G and the QSN. Consequently, the spinor axes and the spinors, defined in the next section, are Dirichlet quantized (normalized).

\subsubsection{Spinors in Dirichlet coordinates}
\label{SD}
For the construction of the icosahedrally symmetric $20G$ we use a family of two successive spinors with $\theta = \frac{2\pi}{q}, \frac{2\pi}{p}$ with axes of $q-$fold and $p-$fold rotations (ref. p. 47 in \cite{Coxeter73}). Here, $q=5 \text{ and } p=3$, as evident by the Schl{\"a}fli notation for the icosahedron, i.e. $\{p,q\} = \{3,5\}$. The origin is located at the center of a cubic section of a FCC lattice that hosts tetrahedrons in eight corners in such a way that all eight tetrahedra share a common vertex at this origin. The starting point of our geometric construction using spinors is one of these eight tetrahedra as will be discussed in the next section. The spinor-axis corresponding to the $\frac{2\pi}{5}$ rotations is given in Dirichlet coordinates as follows:
\begin{align}
{\bf s}^{({5})} &= (0+\phi0){\bf e}_1 + (-1 -\phi3){\bf e}_2 + (2 + \phi1){\bf e}_3 \nonumber \\
&=(0,0){\bf e}_1 + (-1,-3){\bf e}_2 + (2,1){\bf e}_3.
\label{spinor_axis_rot5}
\end{align}
It is important to note that the axis defined above is Dirichlet normalized in the sense that the coefficients $\tilde{s}^{({5})}_{i} = \tilde{s}^{({5})}_{i1} + \phi \tilde{s}^{({5})}_{i2} = (\tilde{s}^{({5})}_{i1},\tilde{s}^{({5})}_{i2}),~i=1,2,3;~\tilde{s}^{({5})}_{i1}, \tilde{s}^{({5})}_{i2} \in \mathbf{Z}$ are Dirichlet numbers. Likewise, the spinor-axis corresponding to the first set of $\frac{2\pi}{3}$ rotations is given in Dirichlet coordinates as follows:
\begin{align}
{\bf s}^{({3})} &= (-1+\phi0){\bf e}_1 + (-1 +\phi0){\bf e}_2 + (1 + \phi0){\bf e}_3 \nonumber \\
&=(-1,0){\bf e}_1 + (-1,0){\bf e}_2 + (1,0){\bf e}_3.
\label{spinor_axis_rot3}
\end{align}
We also note that $$|{\bf s}^{(5)}| = \sqrt{(0 + \phi0)^2 + (-1 -\phi3)^2 + (2 + \phi1)^2},$$ and $$|{\bf s}^{(3)}| = \sqrt{(-1 + \phi0)^2 + (-1 + \phi0)^2 + (1 + \phi0)^2}.$$

The spinors defined in eqs.~\eqref{spinor_def1} and \eqref{spinor_def2} have Dirichlet coefficients, i.e.
\begin{align}
s^{(5)} &= (s^{(5)}_{01} + \phi s^{(5)}_{02}) + (s^{(5)}_{11} + \phi s^{(5)}_{12}) {\bf e}_{23} \nonumber \\ &\quad + (s^{(5)}_{21} + \phi s^{(5)}_{22}) {\bf e}_{31} + (s^{(5)}_{31} + \phi s^{(5)}_{32}){\bf e}_{12} \nonumber \\ 
& \equiv (s^{(5)}_{01},s^{(5)}_{02}) + (s^{(5)}_{11},s^{(5)}_{12}) {\bf e}_{23} + (s^{(5)}_{21},s^{(5)}_{22}) {\bf e}_{31}\nonumber \\ &\quad + (s^{(5)}_{31},s^{(5)}_{32}){\bf e}_{12} \nonumber \\
&\equiv {\bf s}^{(5)}{\bf e}_{123},
\label{spinor_def3}
\end{align} 
To demonstrate this natural occurrence of the Dirichlet coordinates in the expression of the spinor defined by eq \eqref{spinor_def2}, we note that $\theta = \frac{2\pi}{5}, \cos \theta/2 = \phi/2$ where $\phi = \frac{1 + \sqrt{5}}{2}$ is the \textit{golden ratio}. Next, we examine the bivector components of the spinor, i.e., the terms ${\bf e}_{123}\hat{{\bf s}}^{(5)}\sin \frac{\theta}{2}=-\sin  \frac{\theta}{2} \hat{{\bf s}}^{(5)} {\bf e}_{123}$,
\begin{align}
&{\bf e}_{123}\hat{{\bf s}}^{(5)}\sin \frac{\theta}{2} \nonumber \\
&= -\sin \frac{2 \pi}{10} \frac{1}{|{\bf s}^{(5)}|}\biggl\{(\tilde{s}^{(5)}_{11} + \phi \tilde{s}^{(5)}_{12}){\bf e}_{23} \nonumber \\ &\quad \quad + (\tilde{s}^{(5)}_{21} + \phi \tilde{s}^{(5)}_{22}){\bf e}_{31} + (\tilde{s}^{(5)}_{31} + \phi \tilde{s}^{(5)}_{32}){\bf e}_{12}\biggr\} \nonumber \\
&= -\sin \frac{2 \pi}{10} \frac{1}{|{\bf s}^{(5)}|}\biggl\{(\tilde{s}^{(5)}_{11},\tilde{s}^{(5)}_{12}){\bf e}_{23} + (\tilde{s}^{(5)}_{21},\tilde{s}^{(5)}_{22}){\bf e}_{31} \nonumber \\ &\quad \quad + (\tilde{s}^{(5)}_{31},\tilde{s}^{(5)}_{32}){\bf e}_{12}\biggr\} \nonumber \\
&= (0,0){\bf e}_{23} + \biggl(-\frac{1}{2},0\biggr){\bf e}_{31} + \biggl(-\frac{1}{2},\frac{1}{2}\biggr){\bf e}_{12}.
\label{spin_rot5_1}
\end{align}
Thus the Dirichlet normalized spinor corresponding to the $\frac{2\pi}{5}$ rotations is defined as follows,
\begin{align}
s^{(5)} &= 2\biggl(\cos \frac{\theta}{2} + {\bf e}_{123}\hat{{\bf s}}^{(5)}\sin \frac{\theta}{2}\biggr) \nonumber \\
&= (0,1)+(0,0){\bf e}_{23}+(-1,0){\bf e}_{31}+(-1,1){\bf e}_{12},
\label{spinor_rot5}
\end{align}
with the spinor axis defined by eq.~\eqref{spinor_axis_rot5}.
It is noteworthy to compare eqs.~\eqref{spinor_def3} and \eqref{spinor_rot5} which define this spinor in the Dirichlet coordinates.\footnote{The multiplying factor of $2$ in eq.\eqref{spinor_rot5} ensures that the spinor is \textit{Dirichlet} normalized, i.e,. the coefficients in the definition of the spinor are Dirichlet integers (and not made of fractional numbers) of the minimum unit value permissible.} Likewise, the Dirichlet normalized spinor corresponding to the first set of $\frac{2\pi}{3}$ rotations is defined below,
\begin{equation}
s^{(3)} =  (1,0) + (-1,0){\bf e}_{23} + (-1,0){\bf e}_{31} + (1,0){\bf e}_{12},
\label{spinor_rot3}
\end{equation}
with the spinor axis defined by eq.~\eqref{spinor_axis_rot3}. The Dirichlet normalized spinors for the subsequent set of $\frac{2\pi}{3}$ rotations are obtained by successive rotations of ${\bf s}^{(3)}$ by $s^{(5)}$. Summarizing, we use one $s^{(5)}$ spinor and a family of $s^{(3)}$ spinors obtained by successive rotations by $s^{(5)}$ starting with the first $s^{(3)}$ spinor defined by eq.~\eqref{spinor_rot3}.
\subsection{Emergence of the 20G by the action of Clifford motors}
In this section, we demonstrate the action of the spinors ${\bf s}^{(5)}$ and ${\bf s}^{(3)}$ on the vertices of successive tetrahedra starting with a tetrahedron hosted by a cubic section of a FCC lattice which bears the crystallographic root system, $D_3$. A cubic section of a FCC lattice hosts eight different tetrahedra in each of its eight corners. These eight tetrahedra belong to two different families of either right or left handed chiral groups. However, once we pick one of these tetrahedra as the origin of our construction (see Figure (\ref{first_tetrahedron}) below), the chirality of the ensuing collection of tetrahedra remains invariant under the action of the Clifford spinors.   
\begin{figure}[h!]
\begin{center}
\includegraphics[scale=0.23]{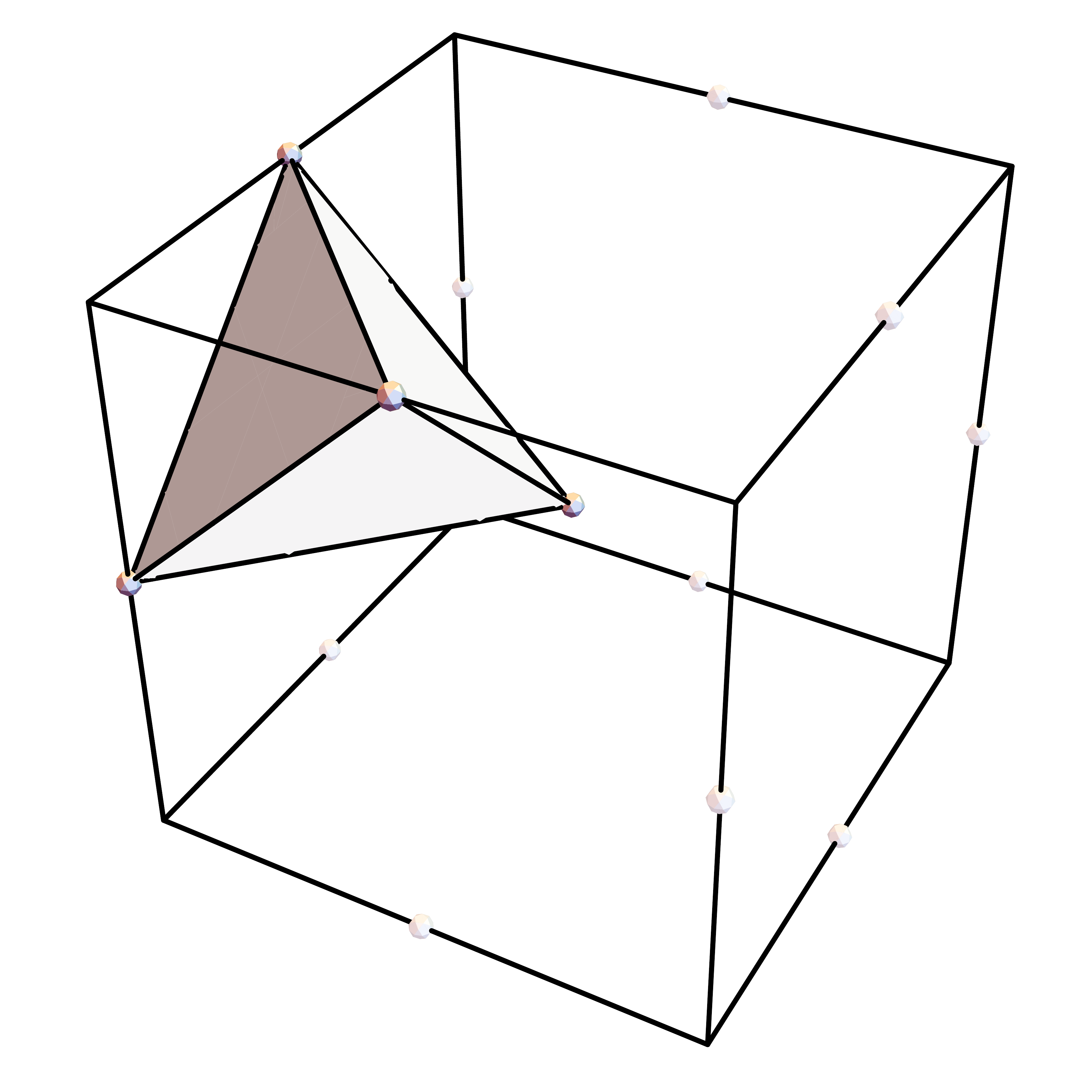}
\end{center}
\caption {The tetrahedron hosted by a cubic section of a FCC lattice is the simplest platonic solid that forms the base of our construction of the icosahedral 20G. The spherical nodes are some of the lattice points of a FCC lattice. The node at the center of the cubic section forms the central vertex (\textit{center}) of the 20G. All tetrahedra that constitute the 20G share one of their vertices at this center.}
\label{first_tetrahedron}
\end{figure}

\subsubsection{The primal 5-group}
\begin{figure*}
  \includegraphics[width=0.245\textwidth]{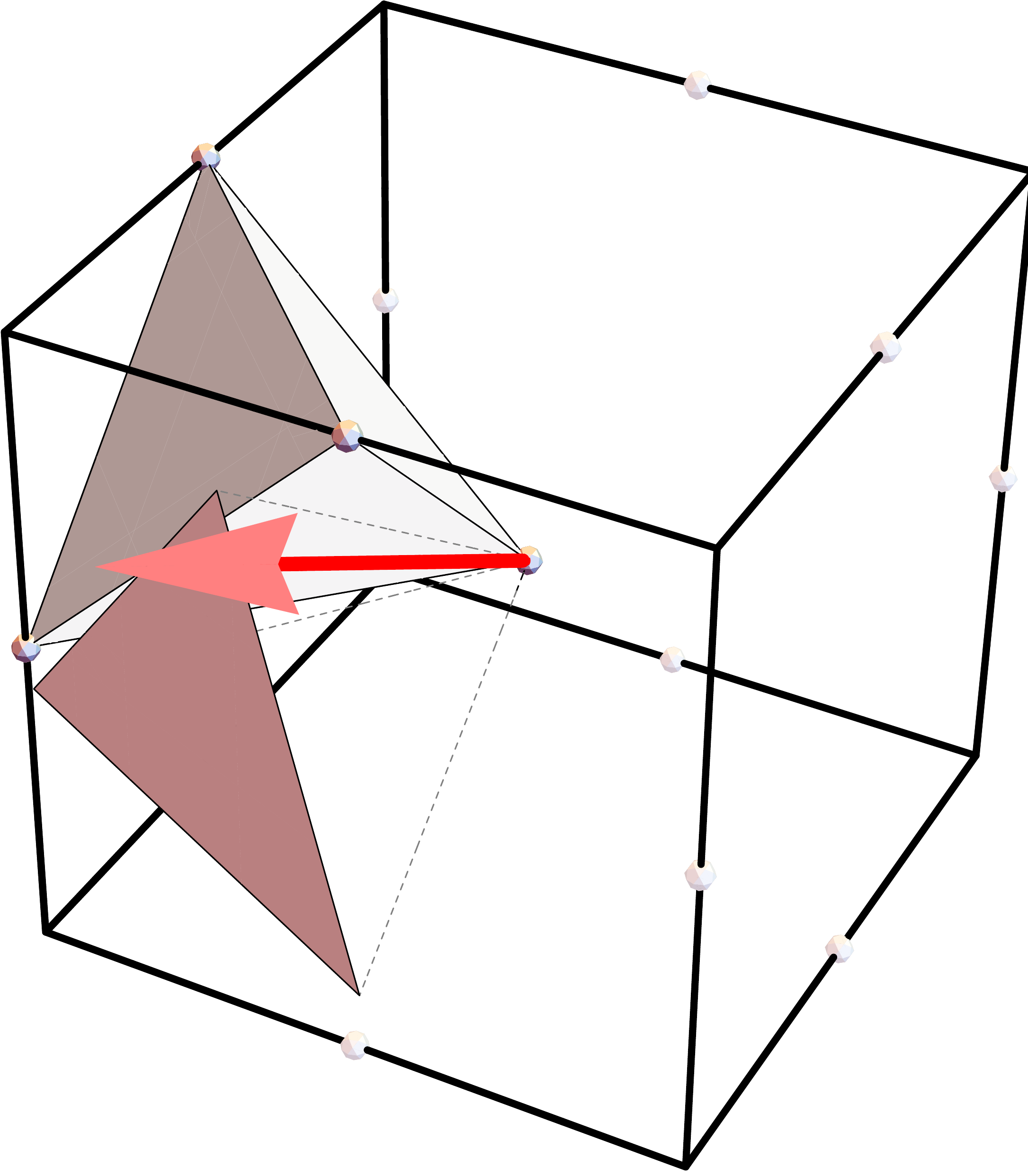}
  \includegraphics[width=0.245\textwidth]{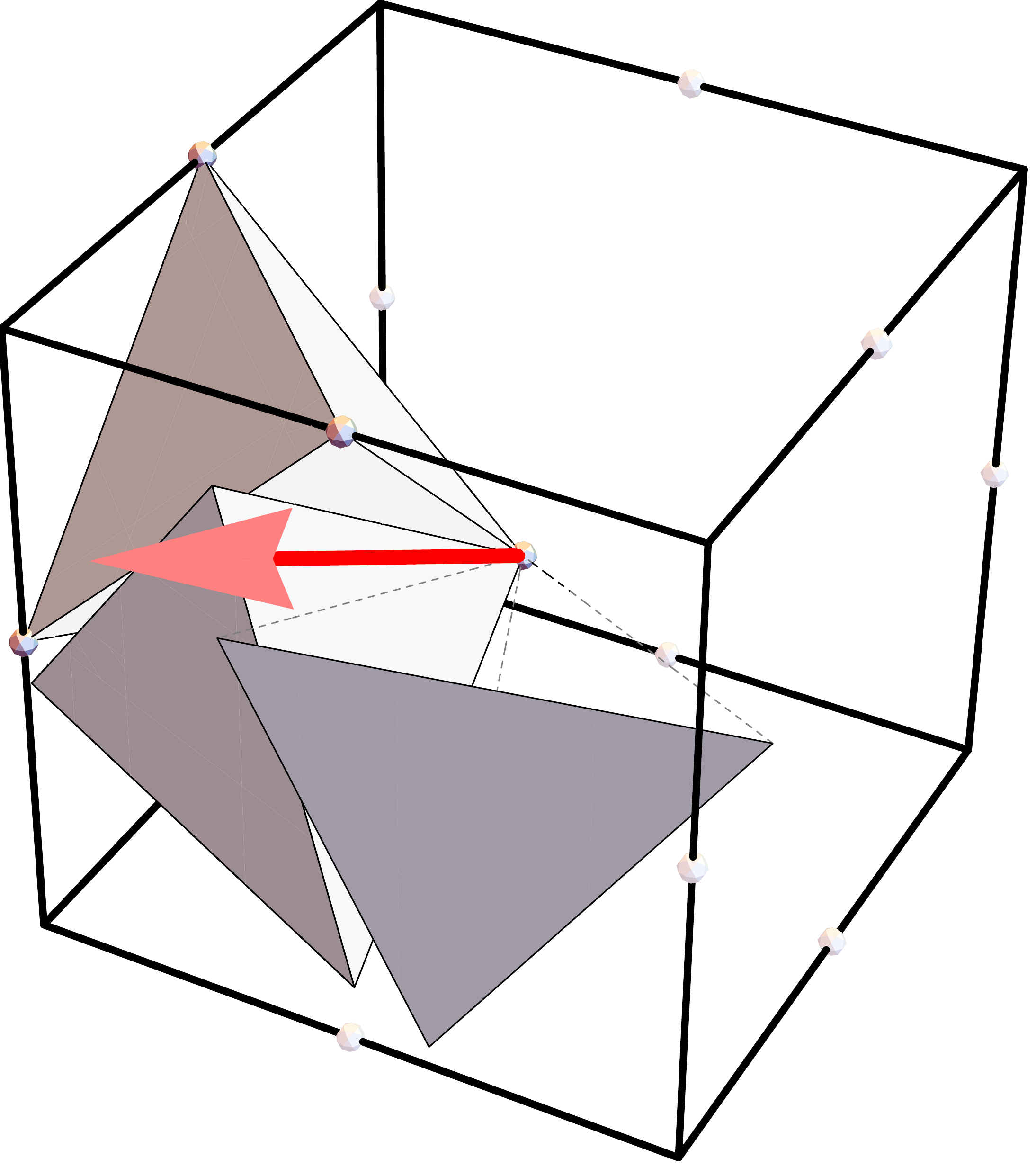}
  \includegraphics[width=0.245\textwidth]{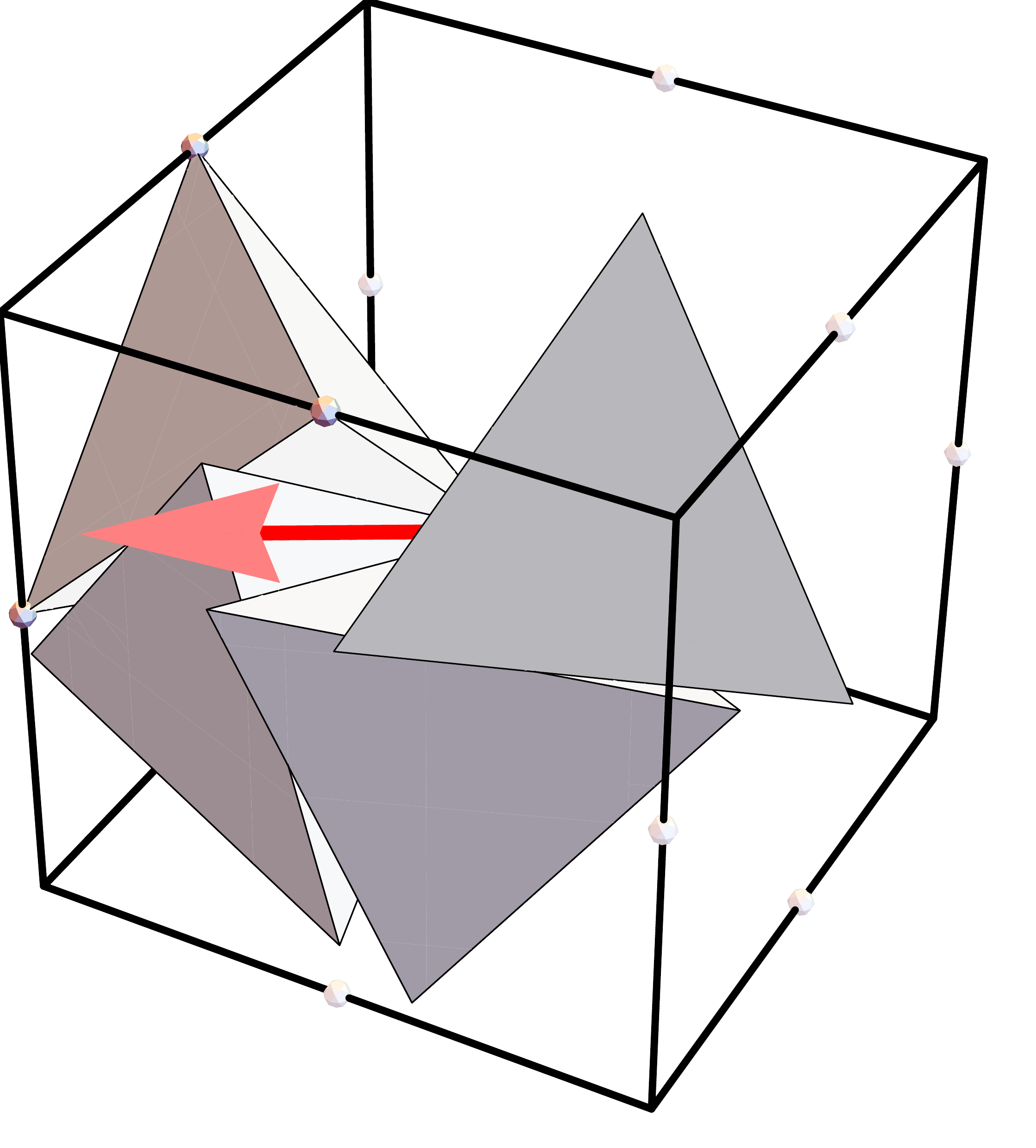}
  \includegraphics[width=0.245\textwidth]{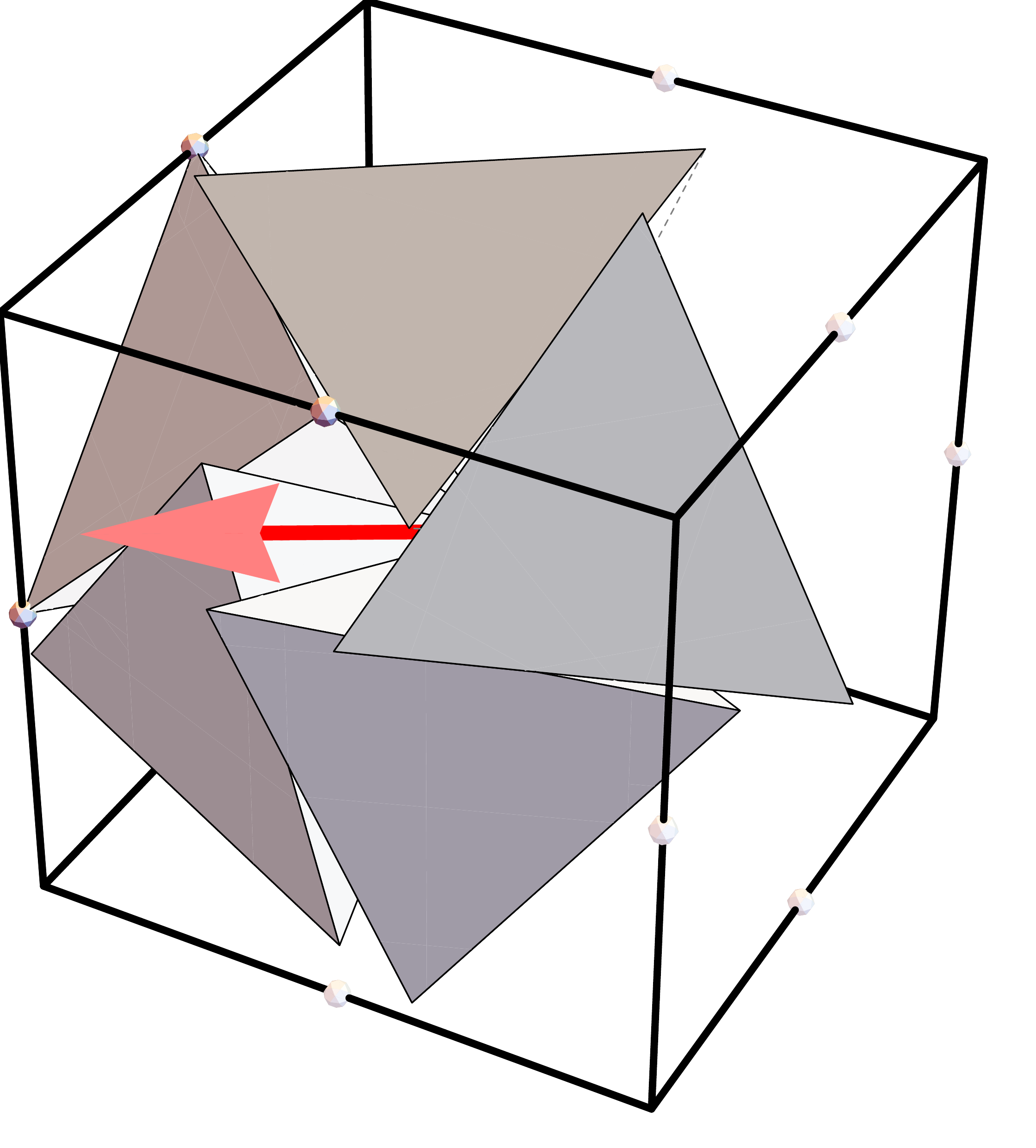}
  \caption{Construction of the first 5G, known as the \textit{primal} 5G, by the action of the spinor $s^{(5)}$ on successively generated tetrahedra. The $s^{(5)}$ spinor is denoted here by the red arrow.}
  \label{primal_5G}
\end{figure*}

In Figure (\ref{primal_5G}), we show the construction of the \textit{primal} 5G\footnote{The \textit{primal} five group (5G) is of fundamental importance as we show subsequently that this 5G is the generator of the remaining tetrahedral clusters, and hence can be regarded as the generator of the 20G by the action of the Clifford motors.} by the action of the spinor $s^{(5)}$ on the successive tetrahedra (specifically, the vertices of the tetrahedra that are position vectors with reference to the central vertex, called the \textit{center}) starting with the originating tetrahedron shown in Figure (\ref{first_tetrahedron}).
The primal 5G (along with the four other 5Gs) are the \textit{canonical} units of the 20G for two reasons.
\begin{enumerate}
\item The primal 5G is the simplest unit of the 20G icosahedral cluster that fixes the chirality of the 20G as has been mentioned earlier.
\item The pentagonal gap associated with the primal 5G (as well as the other 5Gs), as seen in the rightmost graphic in Figure (\ref{primal_5G}) is a direct consequence of the fact that the 20G built from it has the minimal number of plane classes (see discussion in \cite{Fang15}). This makes the 5Gs \textit{canonical} building blocks of the 20G.  
\end{enumerate}

\subsubsection{Generation of the auxiliary 5Gs}
The generation of the remaining 5Gs is described below by the combined action of a family of $s^{(3)}$ spinors followed by the action of a family of Clifford shift vectors upon the primal 5G. This sequence of rotations followed by a set of translations (shifts) together constitute the action of Clifford motors (spinors followed by shifters).\\

\begin{figure*}
  \includegraphics[scale=0.18]{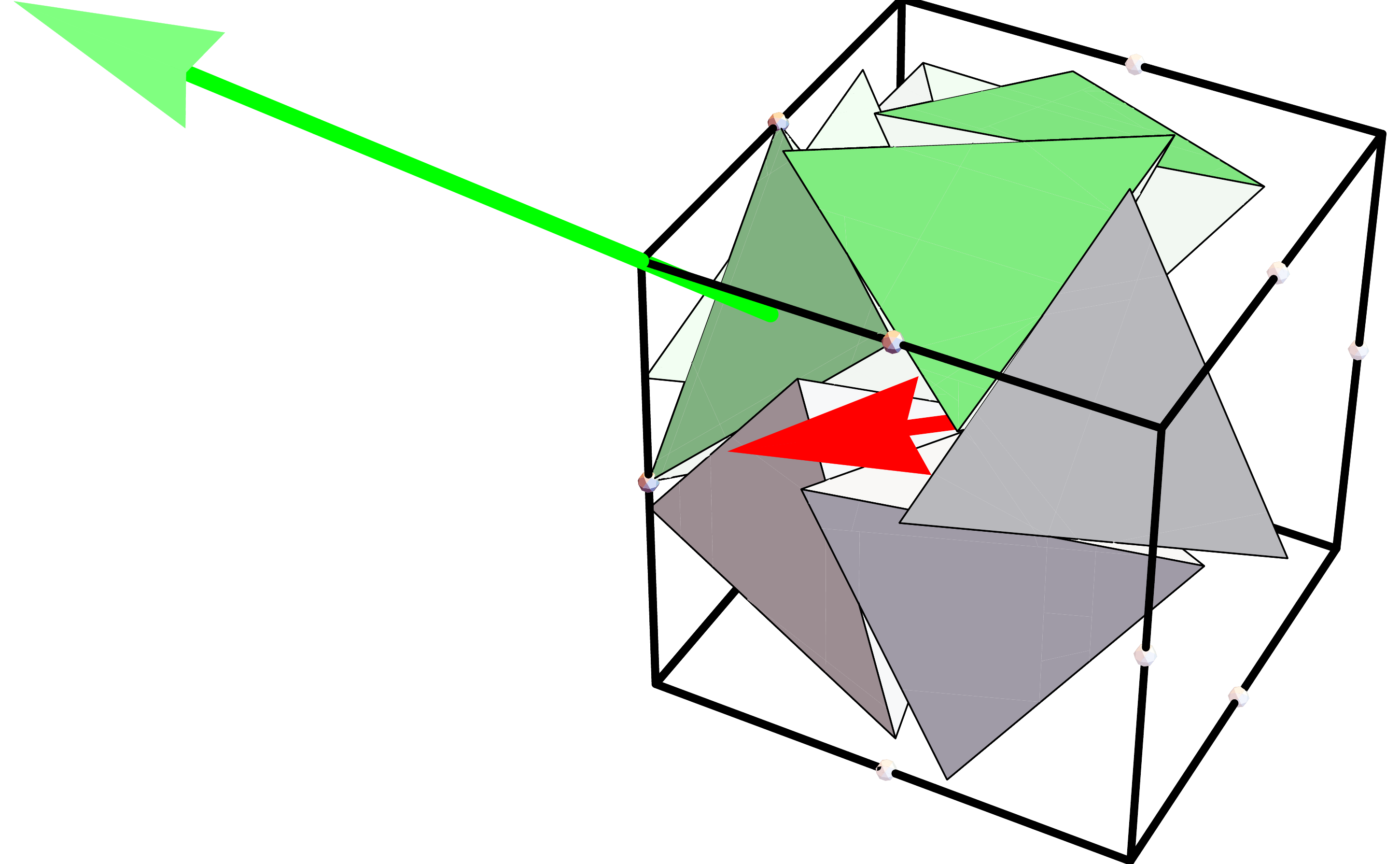}
  \includegraphics[scale=0.18]{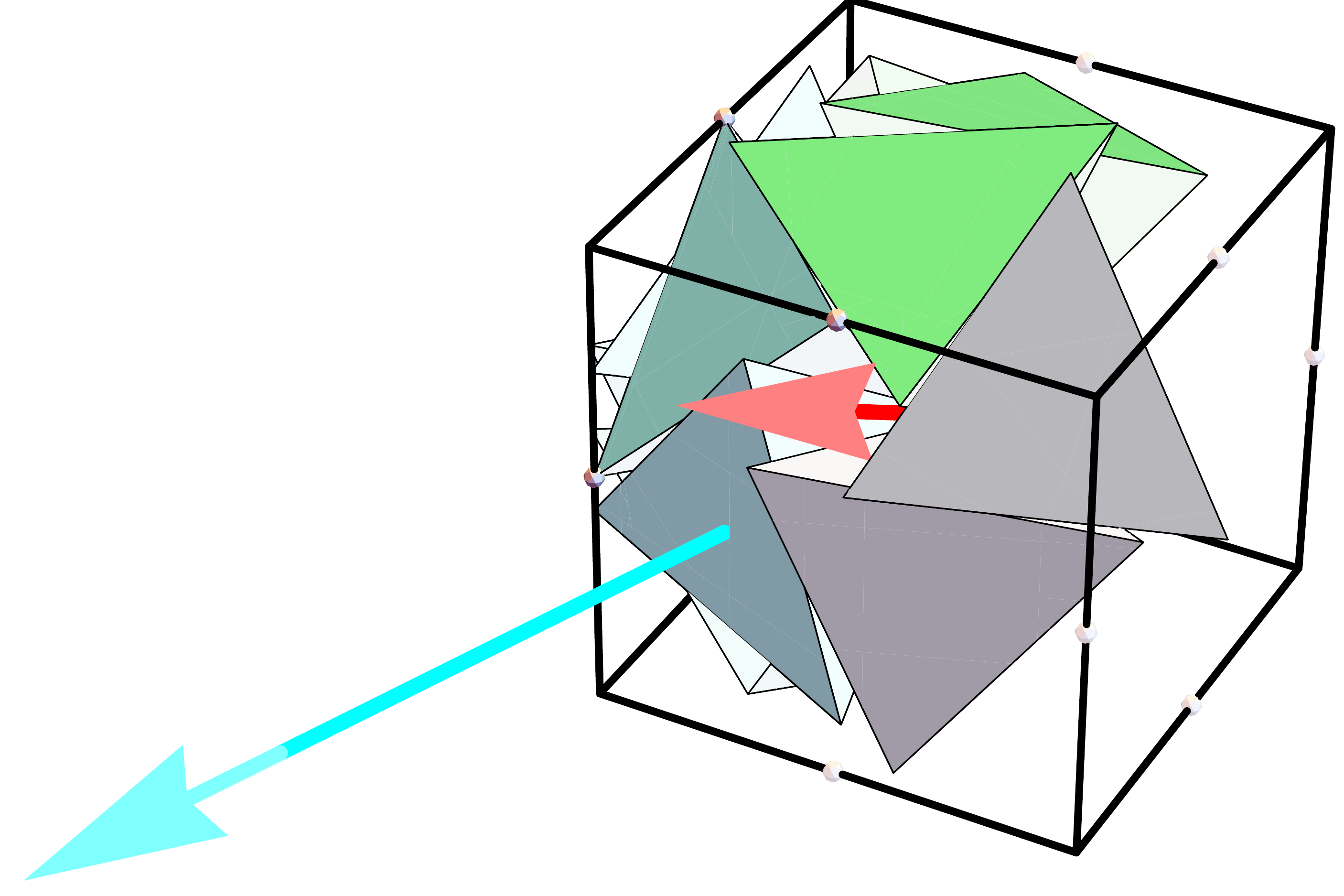}
  \includegraphics[scale=0.16]{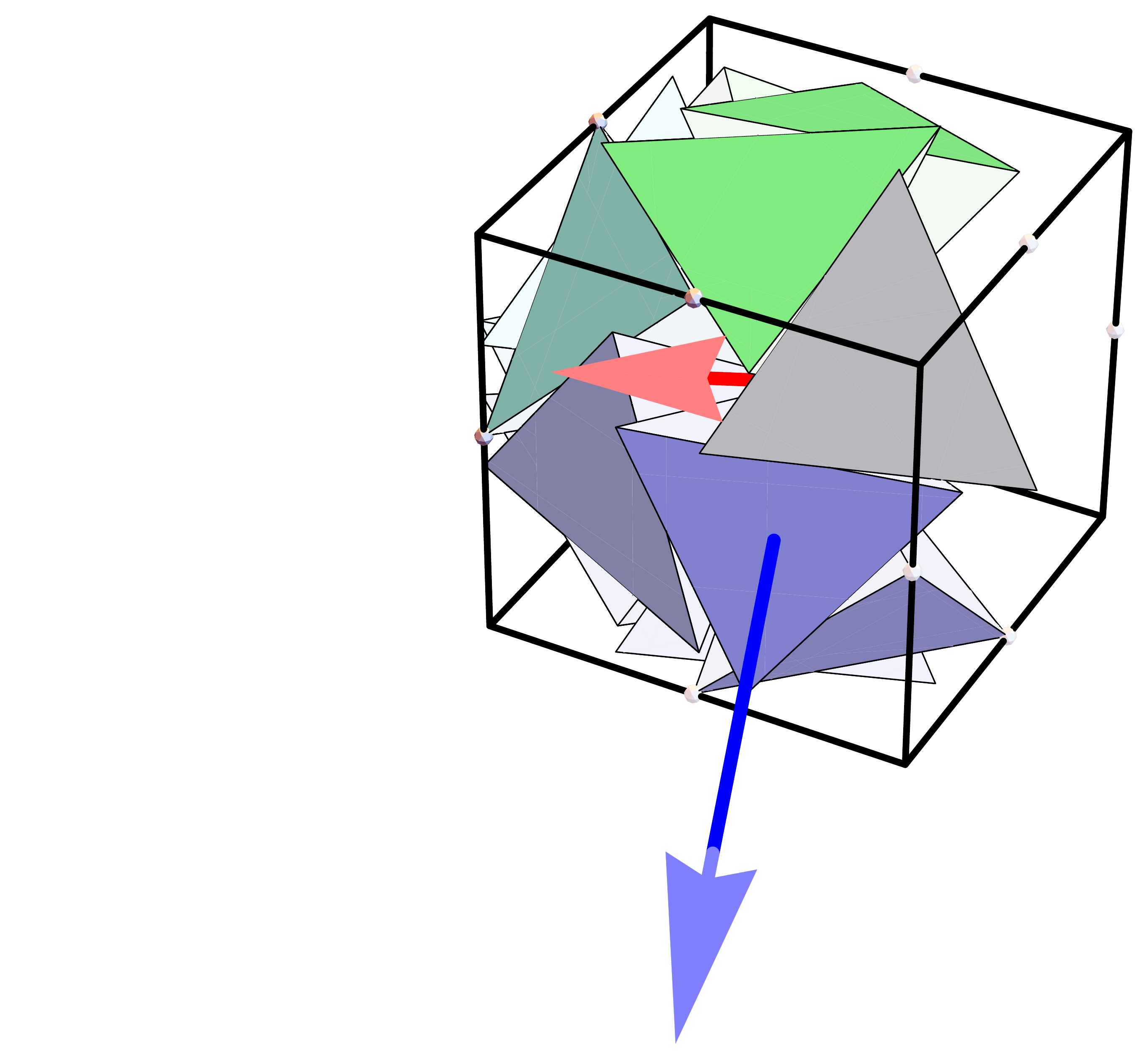}
  \includegraphics[scale=0.19]{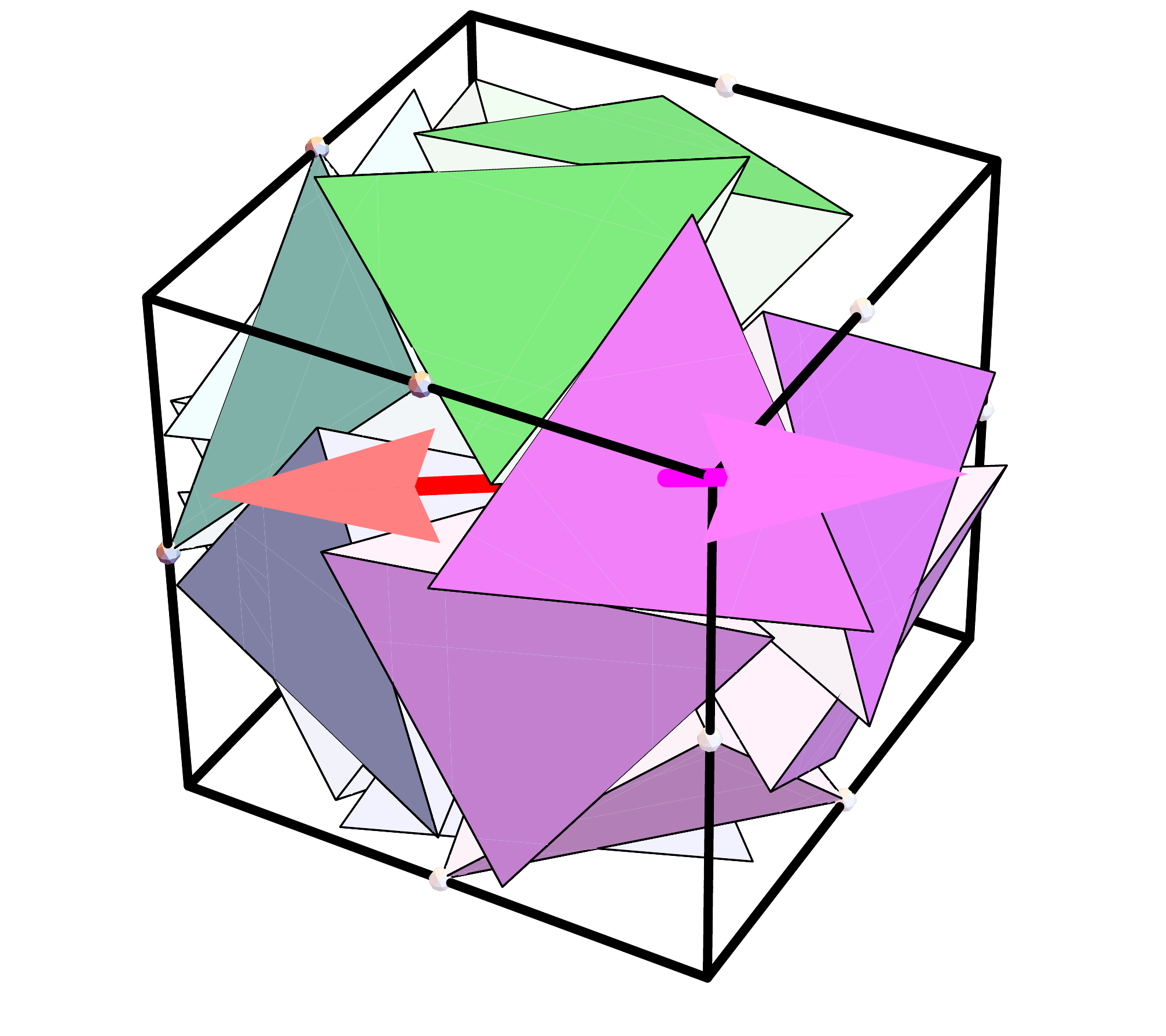}
  \includegraphics[scale=0.23]{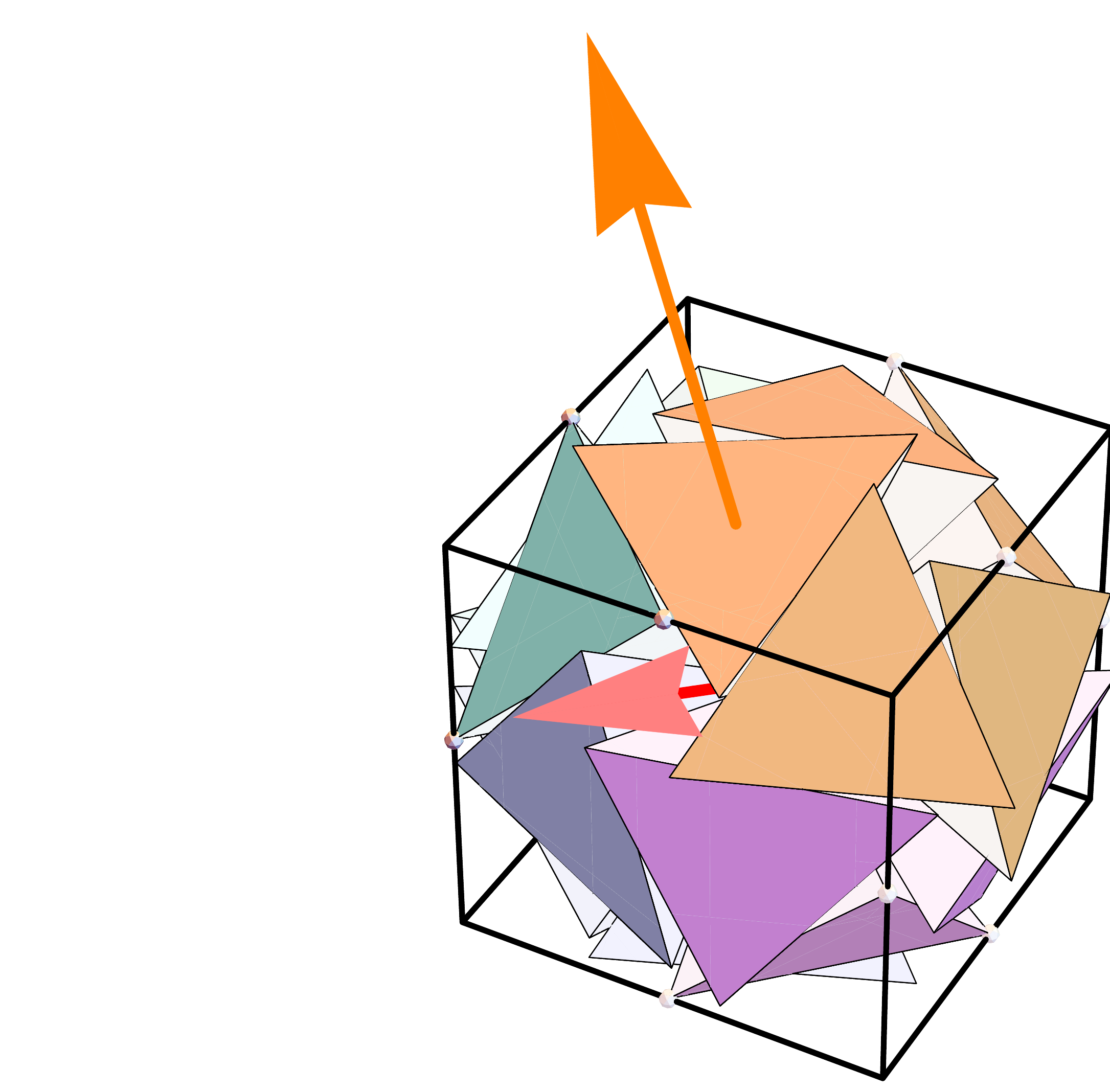}
    \includegraphics[scale=0.21]{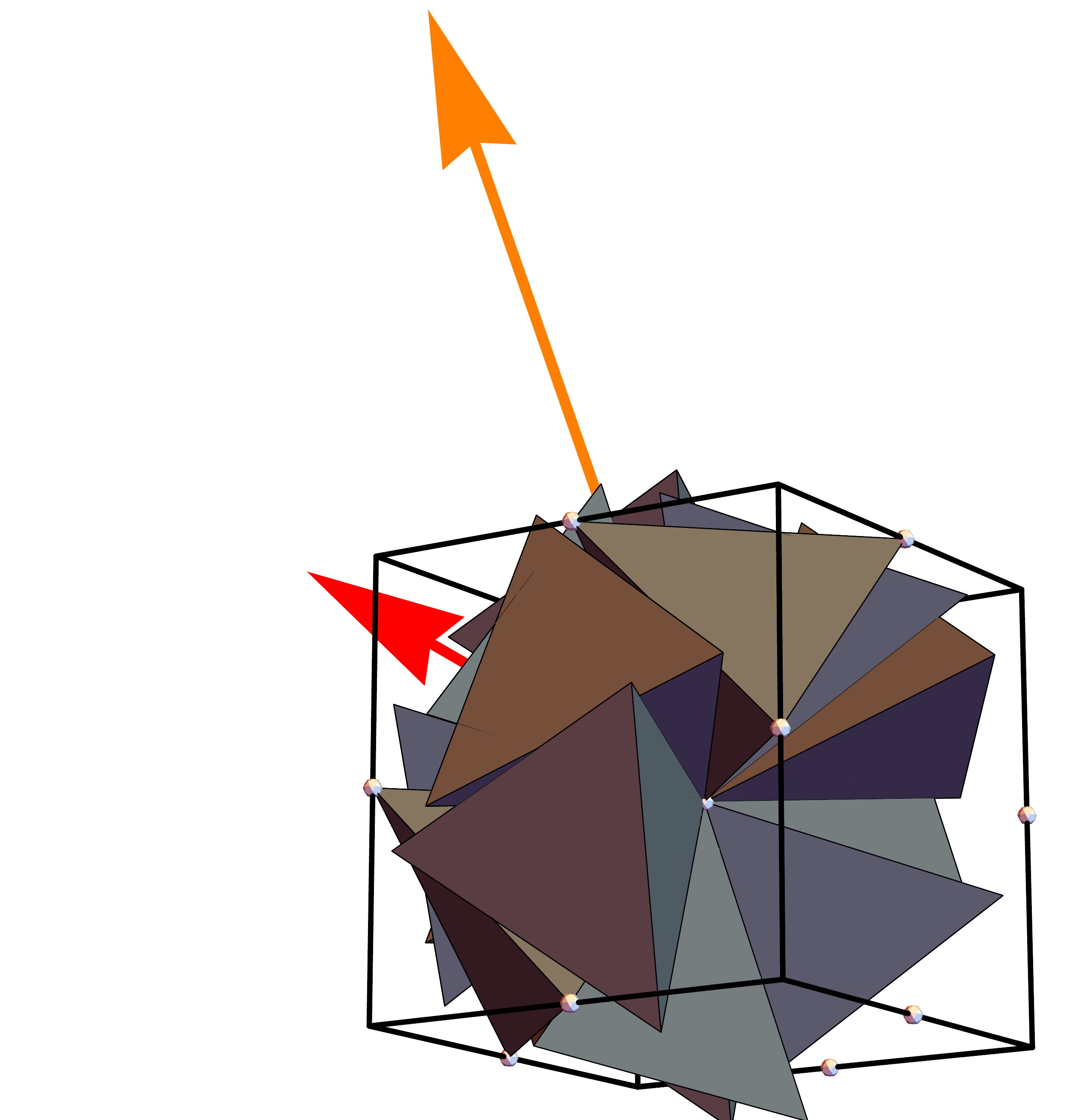}
  \caption{Construction of the auxiliary 5Gs by the action of the family of $s^{(3)}$ spinors on the primal 5G is shown here beginning with top left and moving row wise. The last graphic at the bottom right shows a gap diametrically opposite to the primal 5G. The family of $s^{(3)}$ spinors is designed by successive rotation of the $s^{(3)}$ spinors by the $s^{(5)}$ spinors starting with the first $s^{(3)}$ spinor defined by eq.~\eqref{spinor_rot3}. The $s^{(3)}$ family of spinors is shown here by the non-red arrows.}
  \label{other_5Gs}
\end{figure*}

\noindent \textbf{\underline{Action of $s^{(3)}$ spinors on the primal 5G}}: The primal 5G is rotated by the $s^{(3)}$ spinor given by eq.~\eqref{spinor_rot3}. This means each vertex (and hence their position vectors with respect to the center) of each tetrahedron of the primal 5G is rotated by $s^{(3)}$. The resulting tetrahedral cluster, along with the newly generated 5G shaded in green is shown in the top left graphic of Figure (\ref{other_5Gs}). Consequently, other new sets of 5Gs are generated by the successive action of $s^{(3)}$ spinors which are themselves generated by the rotation of their predecessors by the $s^{(5)}$ spinor as shown in the sequence of graphics in Figure (\ref{other_5Gs}). It must be noted that each generation of 5Gs share some tetrahedral units with its predecessor. 

This process of producing auxiliary 5G clusters may be continued in this manner until production of the 20G. Clearly something more efficient can be done to accomplish this. In the last graphic of Figure (\ref{other_5Gs}), we observe that upon the completion of the sixth generation of 5G clusters, a gap befitting a 5G is created that is diametrically opposite to the primal 5G. This means that a reflection of the primal 5G, by a suitable plane passing through the center and parallel to the plane containing the primal pentagonal gap, followed by an appropriate chirality correction can be used to fill the aforementioned gap and complete the generation of the 20G.\\

\begin{figure*}
  \includegraphics[width=0.33\textwidth]{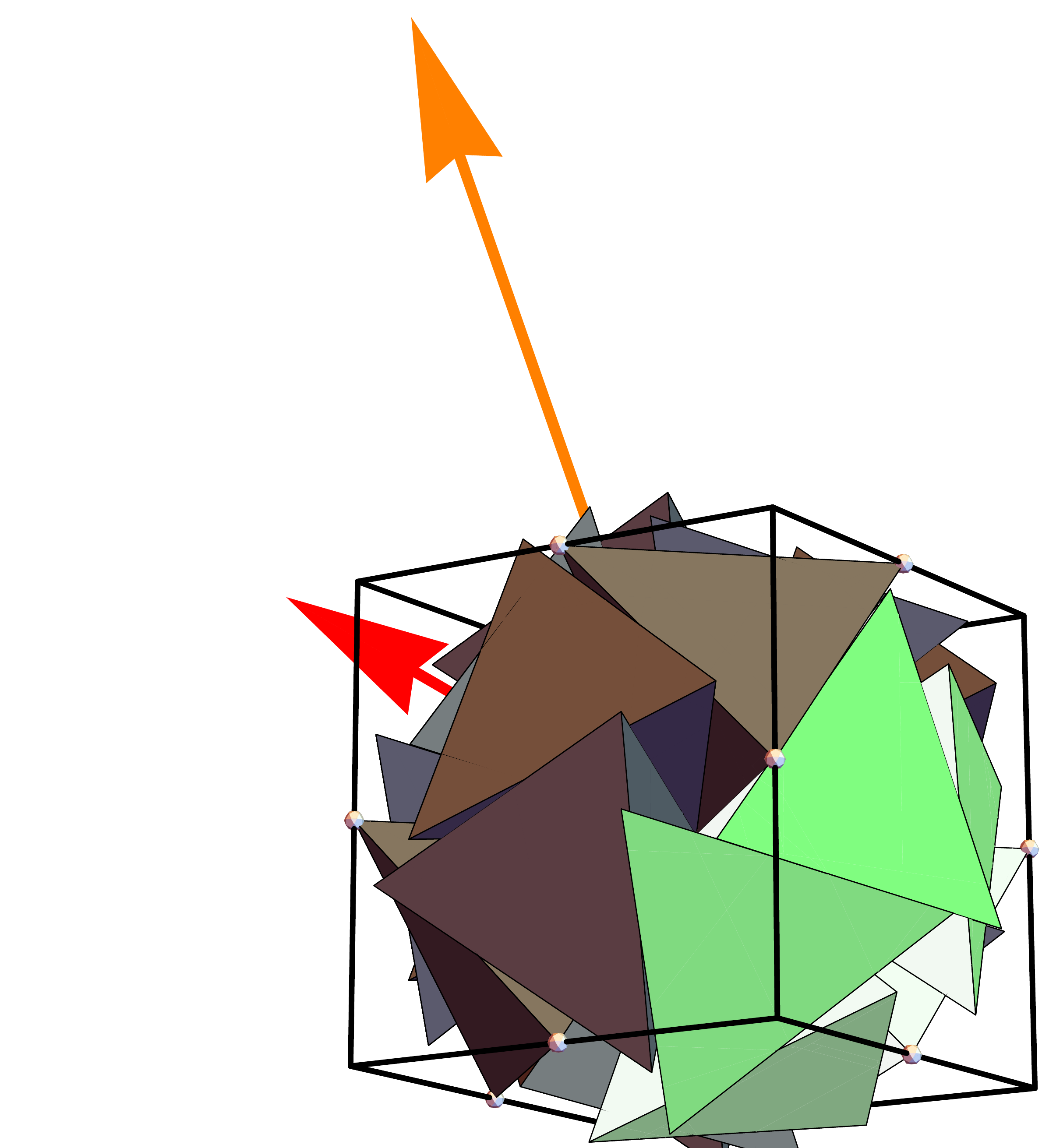}
  \includegraphics[width=0.33\textwidth]{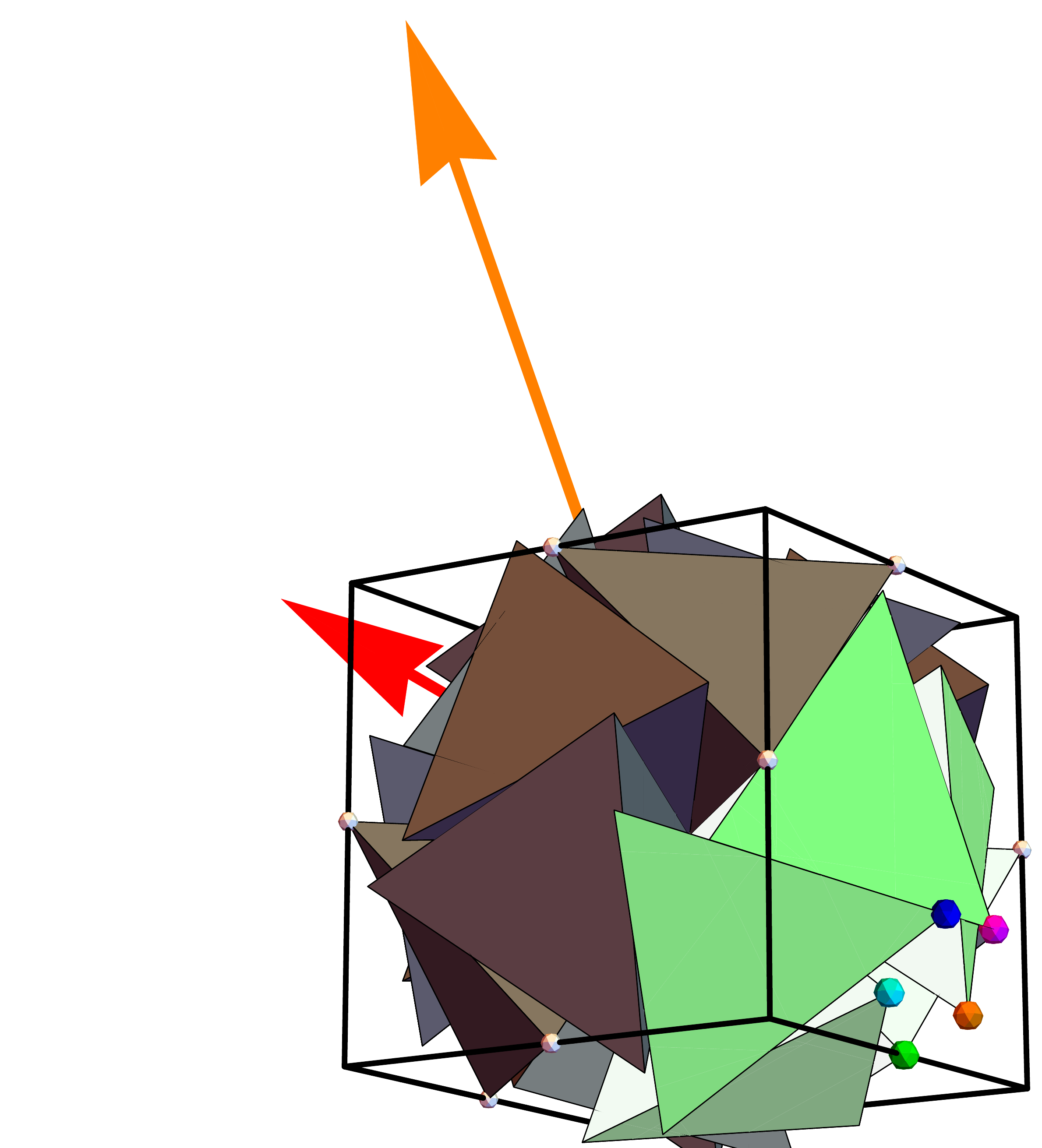}
  \includegraphics[width=0.3\textwidth]{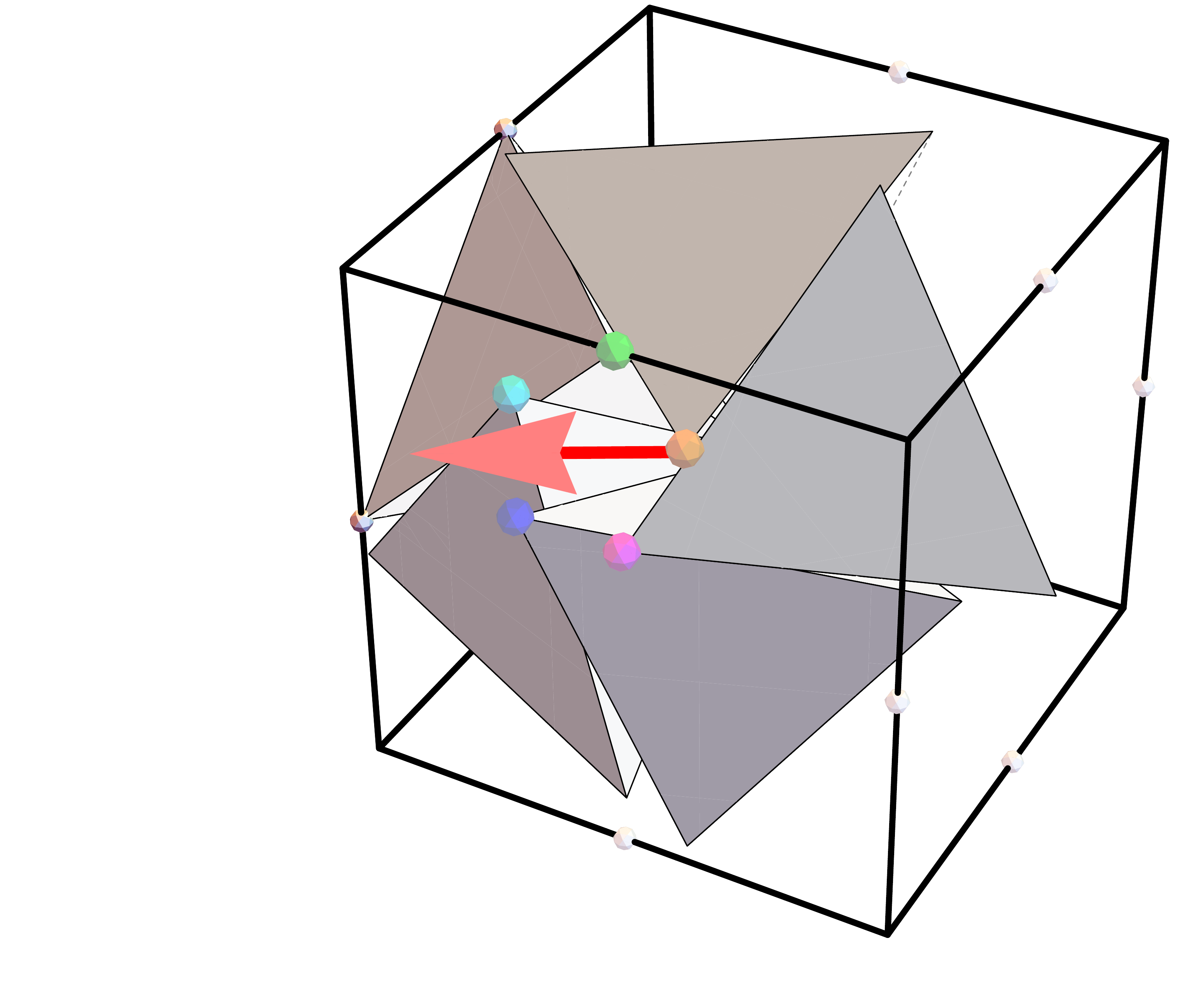}
  \caption{The graphic on the left shows the closure of the gap in the last graphic of Figure (\ref{other_5Gs}) by action of the Clifford shifter on the primal 5G. The middle and right graphics show the correspondence of the vertices of the pentagonal gap of the primal 5G and the newly created 5G. These color matched vertex pairs form the direction axis for the shift operations on the corresponding tetrahedra of the primal 5G.}
  \label{final_5G}
\end{figure*}

\noindent \textbf{\underline{Action of Clifford shifters on the primal 5G}}:
The strategy described in the previous paragraph to fill the gap and complete the generation of the 20G is analogous to a simple shift operation of the primal 5G by a family of Clifford shifters along an appropriate direction. We begin with the first tetrahedron (the one shown in Figure (\ref{first_tetrahedron}) above) of the primal 5G and shift it along the shift vector,
\begin{align}
{\bf n}_{\to} &= (0{\bf e}_1 + 0{\bf e}_2 + 0{\bf e}_3) - (0{\bf e}_1 - 2{\bf e}_2 + 2{\bf e}_3) \nonumber \\
&= 0{\bf e}_1 + 2{\bf e}_2 - 2{\bf e}_3, 
\end{align}
by a magnitude $|{\bf n}_{\to}| = \sqrt{0^2 + 2^2 + (-2)^2} = 2\sqrt{2}$. The direction of the shift vector ${\bf n}_{\to}$ is determined by the edge of the first tetrahedron of the primal 5G that is closest to the $s^{(5)}$ spinor or, equivalently, by the edge containing the center and the vertex that forms the pentagonal gap of the primal 5G. The magnitude of the shift is determined by the distance between the shaded face of the starting tetrahedron shown in Figure (\ref{first_tetrahedron}) and the plane containing the center with a normal vector given by ${\bf n_{\to}}$. This choice of the shift operation guarantees the symmetry of the newly generated final 5G as well as it's chirality. We repeat this shift operation for each of the tetrahedra of the primal 5G. The 20G thus created is shown in Figure (\ref{final_5G}). The 20G has icosahedral symmetry and is the fundamental unit of the aperiodic quasicrystal described in the next section. The vertices of the 20G lie on the Dirichlet coordinate frame and are hosted by two orthogonal $D_3$ lattices, as described in a later section. The interested reader is referred to the article by Fang et. al. \cite{Fang15} for further discussions about the 20G. 

\subsection{Emergence of the Dirichlet quantized quasicrystal}
\begin{figure}
\begin{center}
  \includegraphics[scale=0.33]{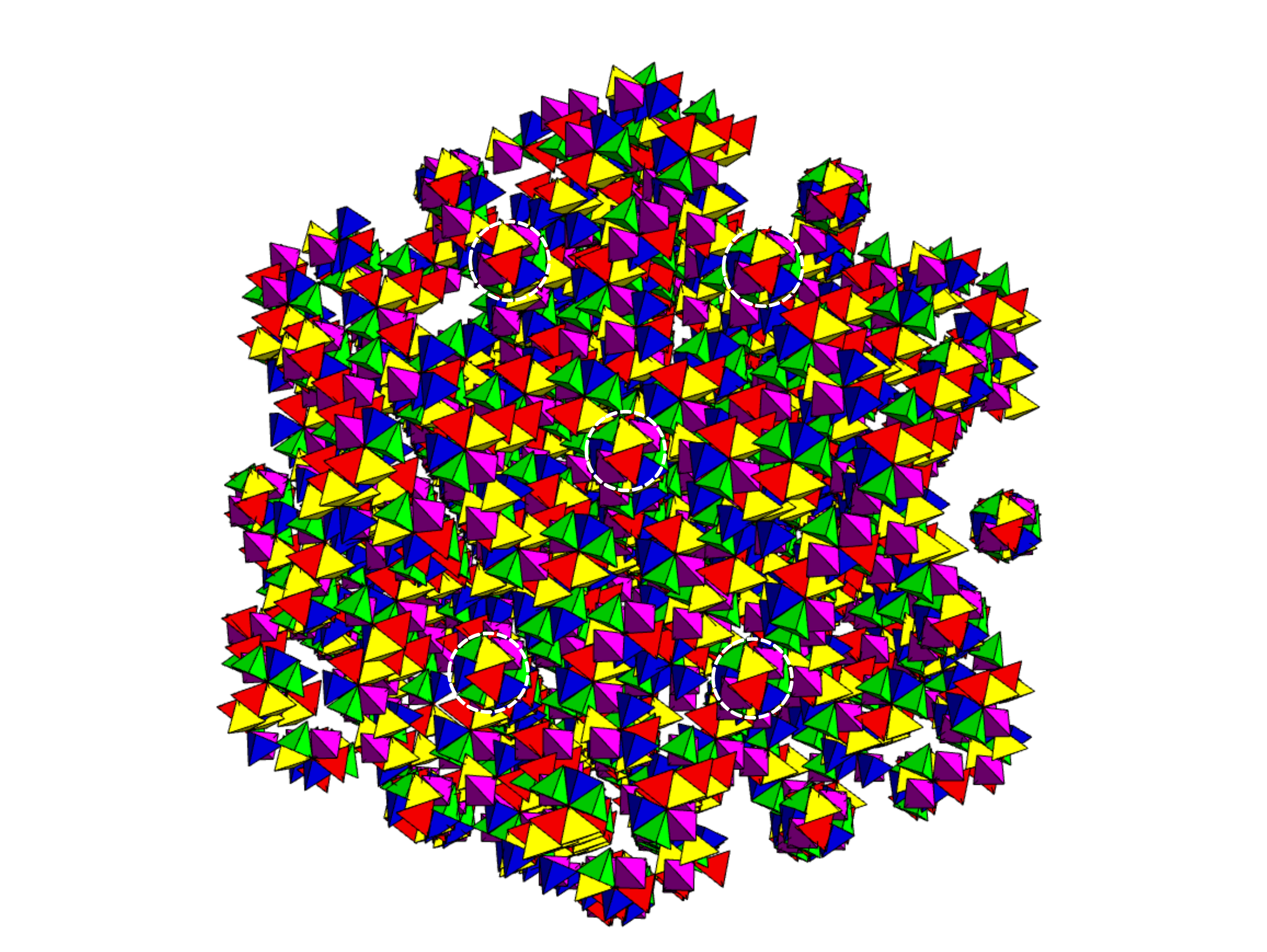}
  \caption{The Dirichlet quantized quasicrystal referred to by Fang et. al. \cite{Fang15} as the QSN.}
  \label{fig13}
  \end{center}
\end{figure}

The 20G does not have aperiodic order and is therefore not a quasicrystal. Aperiodicity is included by spacing the plane sets (classes) of the 20G prescribed by a family of Fibonacci chains made of binary units $1$ and $\phi$ as described in \cite{Fang15}.\footnote{Fang et. al. refer to this Fibonacci spaced quasi-lattice as the \textit{Quasicrystalline Spin Network} (QSN) with the goal of encoding the quantum geometry of space with a suitable language. One of such attempts towards this goal of prescribing a suitable language in terms of Clifford's geometric algebra in Dirichlet space is an underlying objective of this current manuscript.} The resulting object is a quasicrystal ({cf.} Figure \ref{fig13}) and its vertices form a \textit{point set} that also lives in the Dirichlet coordinate frame.\footnote{Since the space of Dirichlet integers is closed under addition and multiplication, the spacing of tetrahedral vertices by $1$ or $\phi$ in the appropriate direction, prescribed by Dirichlet normalized shift vectors, map them to Dirichlet coordinate points.} This can be illustrated mathematically by recalling from sec.~\ref{D} that the units of the Dirichlet quadratic ring are given by $\pm \phi^n$ and noting the following relation \cite{Castro02}:
\begin{equation}
\phi^{n} = (-1)^n F_{n-1} + \phi(-1)^{n+1}F_n,
\label{phi_and_Fib}
\end{equation}
where $\{F_n\}_{n=0,1,2,...}$ are the set of Fibonacci numbers. Eq.~\eqref{phi_and_Fib} clearly shows that the Fibonacci numbers are a special case of Dirichlet numbers and hence the space of vertex points generated by the aforementioned Fibonacci spacing is embedded in the Dirichlet space. This Dirichlet quantized quasi-lattice has a finite set of vertex types  that are catalogued in detail in \cite{Fang15}. The detailed geometric construction and properties of this emergent quasicrystalline structure is also described in \cite{Fang15}.

The topology defined by the vertices of the 20G, along with additional properties such as handedness, vertex types, and the quasicrystalline diffraction pattern of the quasicrystal (see Fang et. al. \cite{Fang15}) invokes further understanding of the emergent geometry that is hosted by the Dirichlet coordinate frame. We present a detailed exploration of the emergent Dirichlet geometry, accompanied by a two-fold spawning of dimensions, in the following section. 

\section{Emergence of the 6D Dirichlet quantized host, $\Lambda_{D_3} \bigoplus \phi \Lambda_{D_3} $}
\begin{figure*}
\begin{center}
\includegraphics[scale=0.45]{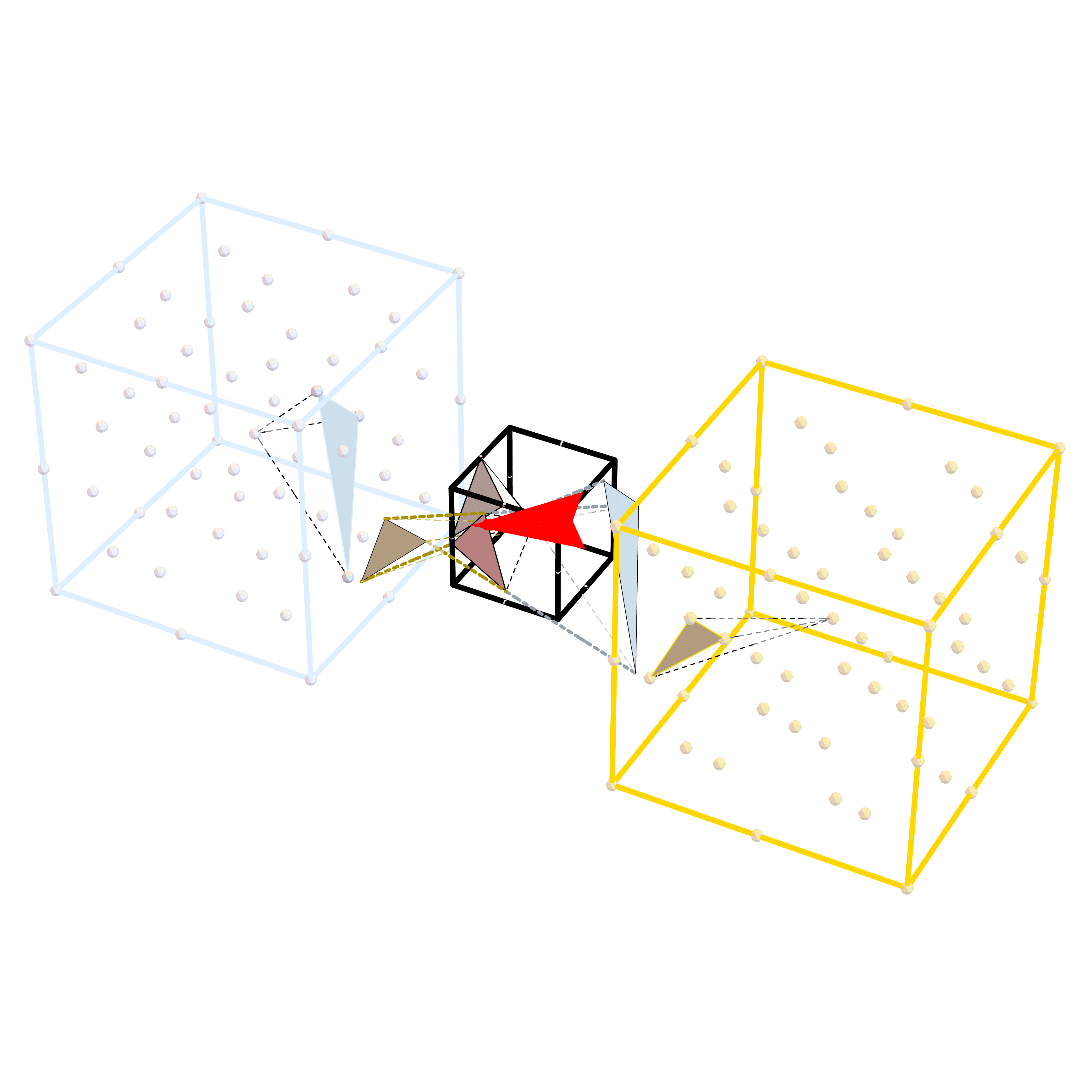}
\end{center}
\caption {The graphic depicts the doubling of dimensions by the action of the Dirichlet quantized spinor, where successively generated tetrahedra have vertices with Dirichlet coordinates with non-golden and golden parts, each being hosted by the three dimensional crystallographic root system $D_3$. In this figure, the newly generated tetrahedron has a golden (colored brown) and non-golden component (colored cyan). Each of their vertices are at a distance $\frac{1}{\phi}$ and $1$ units from the corresponding vertices of tetrahedra hosted by the original FCC lattice (middle) respectively. For the sake of visual clarity, the lattice hosting these golden and non-golden components are translated in space and shown on the sides of the original lattice.}
\label{Emerging_D3xD3}
\end{figure*}

\begin{figure*}
\begin{center}
\includegraphics[scale=0.45]{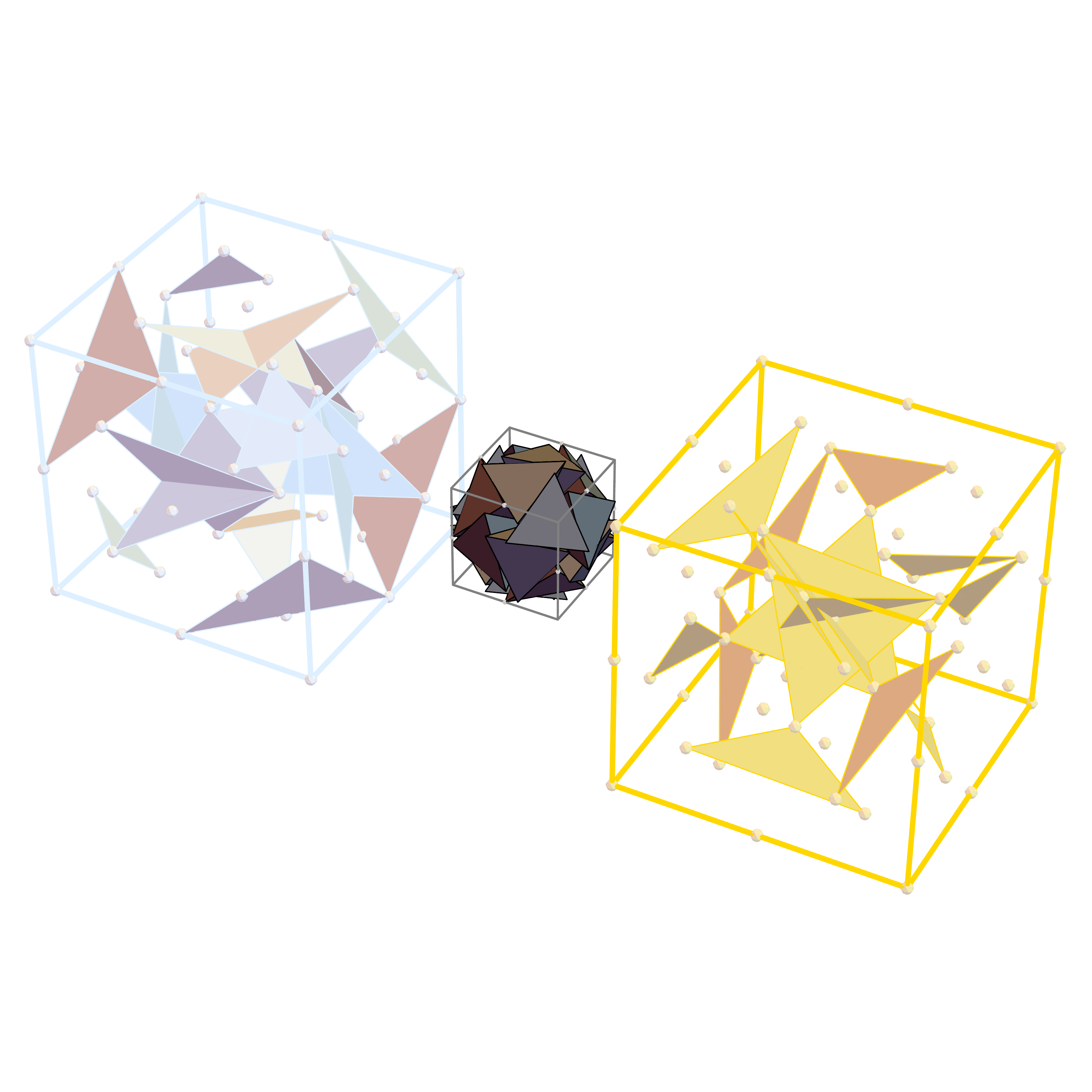}
\end{center}
\caption {The six dimensional Dirchlet quantized lattice hosting the 20G. In other words, the geometric fabric of the icosahedral object (20G) is unraveled and presented as a six dimensional host $\Lambda_{D_3} \bigoplus \phi \Lambda_{D_3} $.}
\label{Dirichlet_host_full}
\end{figure*}

\begin{figure*}
\begin{center}
\includegraphics[scale=0.3]{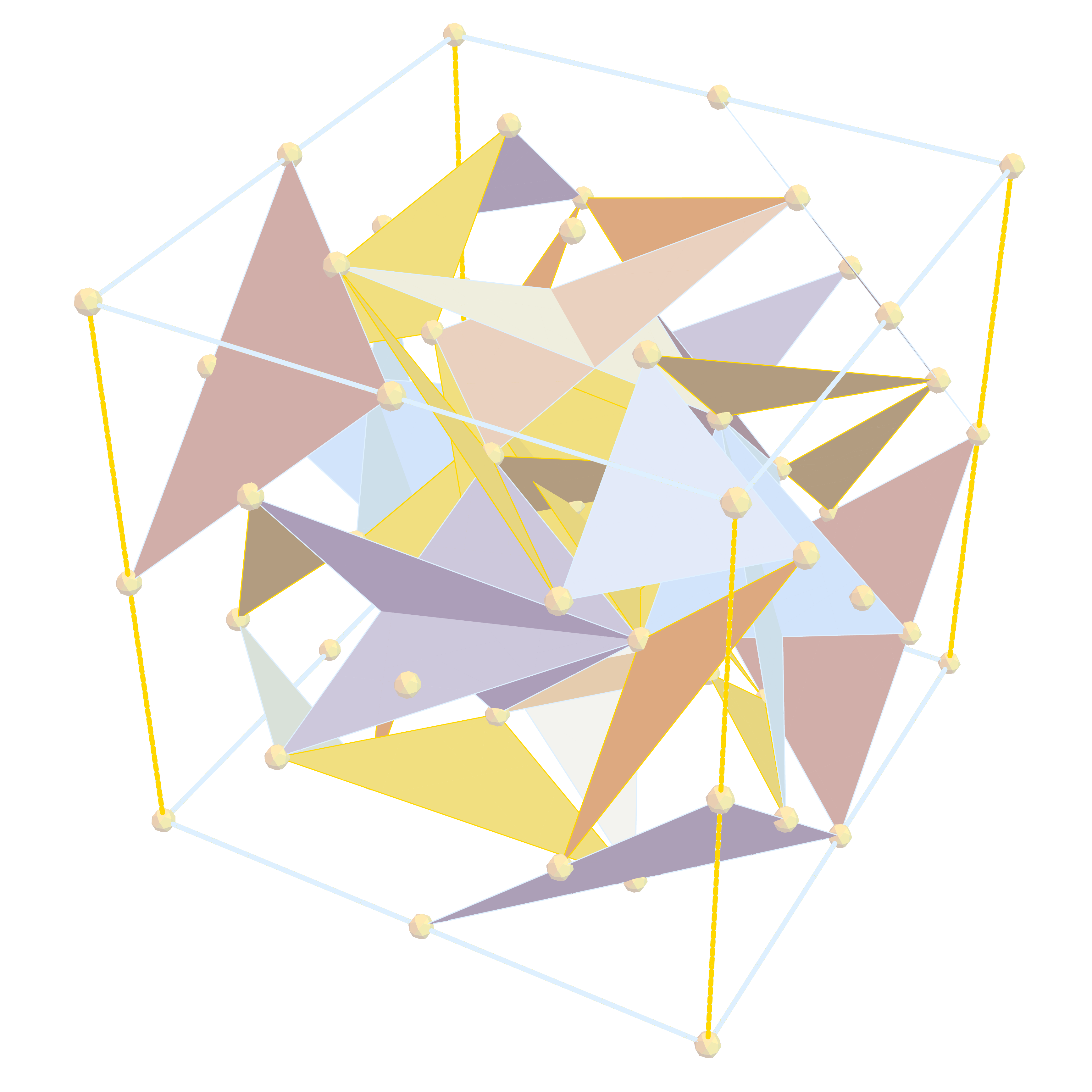}
\end{center}
\caption {Superposition of the golden and non-golden components of the Dirichlet coordinate frames revealing the \textit{orthogonal} relationship between them.}
\label{Dirichlet_OG}
\end{figure*}

Recall that the vertices of the 20G (and the QSN) are Dirichlet coordinates, i.e. they are of the form $(x+\phi x', y+\phi y', z+\phi z')$ where $x,x',y,y',z,z' \in \mathbf{Z}$. This is by construction true as described in sections \ref{D} and \ref{SD} above and inherently linked to the icosahedral symmetry of the 20G. Subsequently, the Dirichlet coordinates are written as ordered pairs of the form $\{(x,y,z), (x',y',z')\}$ resulting in a collection of 6D coordinates (two sets of orthogonal 3D coordinates). The pairing is comprised of a non-golden part and a golden part,\footnote{Non-golden part of $a+\phi b$ is $a$ and golden part is $b$.} each embedded in one of two orthogonal 3D coordinate frames respectively. This is illustrated in detail in Figures (\ref{Emerging_D3xD3}), (\ref{Dirichlet_host_full}) and (\ref{Dirichlet_OG}).

The mapping between the 20G (vertices) that is embedded in the original lattice and the inductively generated component lattices embedded in the $\Lambda_{D_3} \bigoplus \phi \Lambda_{D_3} $ root system is \textit{bijective}, as the vertex pair $(p, p')$ uniquely associates with the corresponding vertex $(p + \phi p')$ of the 20G in the original lattice. It is important to note that the resultant six dimensional structure comprising the golden and non-golden elements of the tetrahedral vertices is \textit{embedded} in the $\Lambda_{D_3} \bigoplus \phi \Lambda_{D_3} $ root system but does not entirely fill the lattice space of $\Lambda_{D_3} \bigoplus \phi \Lambda_{D_3} $. The {embedding} is simple\footnote{The embedding is the marginal identity map $p \star \mathbb{1}_{p  \in \mathbf{D}}, \text{ where }p \star 1 = p, ~p\star 0 = \{ \}$ that maps all points in the embedded space to itself. $\mathbf{D}$ is the embedded space with Dirichlet coordinates.} and is topologically invariant of $\Lambda_{D_3} \bigoplus \phi \Lambda_{D_3} $. We refer to $\Lambda_{D_3} \bigoplus \phi \Lambda_{D_3} $ as the Dirichlet quantized host.\footnote{Here the word \textit{host} is used to emphasize the important fact that the 6D $\Lambda_{D_3} \bigoplus \phi \Lambda_{D_3} $ \textit{hosts} the 3D quasicrystalline substrate as depicted pictorially in Figure (\ref{Dirichlet_host_full}).} 

\subsection{Vertices of 20G as embeddings in $\Lambda_{D_3} \bigoplus \phi \Lambda_{D_3} $ lattice}
We present a brute force proof that the vertices of the 20G are embeddings in the $D_3 \times D_3$ lattice. The geometry of the $D_3$ lattice is such that the sum of the coordinate points of a $D_3$ lattice is even \cite{Sirag16}. This requirement demands that the vertices of the 20G with coordinates $(x + \phi x', y + \phi y', z + \phi z')$ must satisfy the following criteria:
\begin{align}
&(x + y + z) \% 2 = 0, \text{ and } \nonumber \\
&(x' + y' + z') \% 2 = 0,
\label{20G_are_D3} 
\end{align}
where $\%$ denotes the modulo operation. 

In the Appendix, we provide data of all the coordinate points of the vertices of the 20G where it can be easily verified that the conditions listed in eq. (\ref{20G_are_D3}) are satisfied. A more elegant proof will require a better understanding of why the action of spinors on the tetrahedra results in new tetrahedra with vertices satisfying the above conditions and will entail rigorous investigation using tools from the theory of algebraic rings and combinatoric approaches. This is beyond the scope of the current manuscript.   

In an identical manner, it can be verified that the coordinates of the QSN are embedded in the $\Lambda_{D_3} \bigoplus \phi \Lambda_{D_3} $ lattice. It is also essential to emphasize that the vertices of the 20G and the QSN are {embeddings} in the $\Lambda_{D_3} \bigoplus \phi \Lambda_{D_3} $ lattice. This means not all lattice points of $\Lambda_{D_3} \bigoplus \phi \Lambda_{D_3} $ are vertices of the 20G and the QSN. Moreover, while the QSN is \textit{aperiodic} in its point space, the golden and the non-golden parts of its vertex coordinates are each embedded in a periodic $D_3$ lattice.  


\subsection{Representation of $\Lambda_{D_3} \bigoplus \phi \Lambda_{D_3} $ and its root system}
\begin{figure}[h!]
\begin{center}
\includegraphics[scale=0.5]{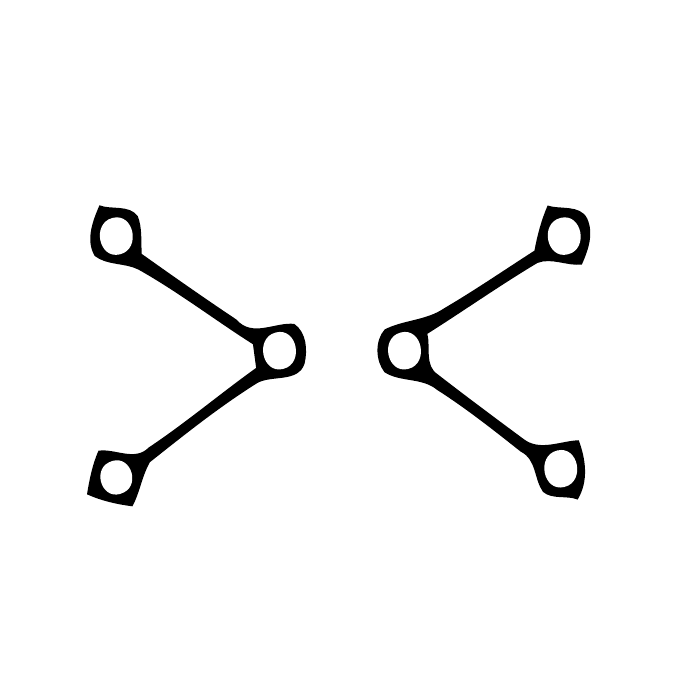}
\end{center}
\caption {Dynkin graph of $\Lambda_{D_3} \bigoplus \phi \Lambda_{D_3} $}
\label{D3xD3}
\end{figure}

The Dirichlet quantized host, $\Lambda_{D_3} \bigoplus \phi \Lambda_{D_3} $ has the following representation in Lie theory using Dynkin graphs \cite{Hump90, Fuchs97, Das14} as shown in Figure (\ref{D3xD3}). The Dynkin diagram is clearly representative of the fact that the 6D Dirichlet host lattice is comprised of two independent $D_3$ component lattices. 
Consequently, the Cartan matrix \cite{Hump90, Fuchs97, Das14} can be deduced from the Dynkin graph shown in Figure (\ref{D3xD3}) as follows:  
\begin{eqnarray}
&&\bigl[ \Lambda_{D_3} \bigoplus \phi \Lambda_{D_3} \bigr] \nonumber \\ &&=\begin{pmatrix}
2 & -1 & -1 & 0 & 0 & 0\\
-1 & 2 & 0 & 0 & 0 & 0\\
-1 & 0 & 2 & 0 & 0 & 0\\
0 & 0 & 0 & 2 & -1 & -1 \\
0 & 0 & 0 & -1 & 2 & 0\\
0 & 0 & 0 & -1 & 0 & 2\\  
\end{pmatrix}, \nonumber
\end{eqnarray}
which can be easily recognized as a block diagonal form of two $D_3$ Cartan matrices. The roots of $\Lambda_{D_3} \bigoplus \phi \Lambda_{D_3} $ can be easily computed from the Cartan matrix \cite{Georgi99}.\footnote{Any set of roots are not unique. The ones mentioned here are the ones that can be easily computed from the Cartan matrix.} The simple roots can be read off from the rows of the matrix $\bigl[ \Lambda_{D_3} \bigoplus \phi \Lambda_{D_3} \bigr]$ and are listed below, 
$$
\alpha_1 = \begin{pmatrix}2\\ -1 \\ -1 \\ 0 \\ 0 \\ 0 \\\end{pmatrix}, \alpha_2 = \begin{pmatrix} -1\\ 2\\ 0\\ 0\\ 0\\ 0\\\end{pmatrix}, 
\alpha_3 = \begin{pmatrix} -1\\ 0\\ 2\\ 0\\ 0\\ 0\\\end{pmatrix},
$$
$$
\alpha_4 = \begin{pmatrix} 0\\ 0\\ 0\\ 2\\ -1\\ -1\\ \end{pmatrix},
\alpha_5 = \begin{pmatrix} 0 \\ 0 \\ 0 \\ -1 \\ 2\\ 0\\  \end{pmatrix}, 
\alpha_6 = \begin{pmatrix} 0 \\ 0 \\ 0 \\ -1 \\ 0\\ 2\\  \end{pmatrix}. 
$$
The remaining roots are $\alpha_1 + \alpha_2, \alpha_1 + \alpha_3, \alpha_1 + \alpha_2 + \alpha_3, \alpha_4 + \alpha_5, \alpha_4 + \alpha_6, \alpha_4 + \alpha_5 + \alpha_6$ and the negative copies of each of the roots listed above. In total, there are $24$ roots of $\Lambda_{D_3} \bigoplus \phi \Lambda_{D_3} $.  In subsequent sections, we present the action of a sequence of bounded linear operators (transformations) on the Cartan matrix of $\Lambda_{D_3} \bigoplus \phi \Lambda_{D_3}  \times A_1 \times A_1$ and the concomitant changes in the topology of the Dynkin graphs and the corresponding root system. We show that the response to this action (sequel) spans the root system of higher dimensional space that embed different unification frameworks of physical matter and forces, viz., $SU(5), E_6$ and $E_8$.   

\section{Emergence of $SU(5), E_6$ and $E_8$ from $\Lambda_{D_3} \bigoplus \phi \Lambda_{D_3} $}
We begin by presenting a flow chart (ref. Figure (\ref{D6toE8}) below) depicting the emergence of a 6D Dirichlet quantized host. The isolated green nodes associated with the Dynkin graphs for each of the intermediary lattices, starting with that of the Dirichlet host and except for the graph of $E_8$, represents the hidden time dimension of curved spacetime. The action of the Dirichlet spinors and the transformation matrices $T_i$ shown in Figure (\ref{D6toE8}) is invariant in time. 
\subsection{Transformations that encode emergence of spacetime fabric}
The emergence of $E_8$ is brought about by the action of a sequence of transformations on the respective Cartan matrices as follows:
\begin{align}
&\breve{T}_2 \circ T_1 \circ \bigl [D_3 \bigoplus D_2\bigoplus A_1 \bigoplus 6A_1 \bigr] = E_4, \nonumber \\
&T_2 \circ T_1 \circ \bigl [D_3 \bigoplus D_2\bigoplus A_1 \bigoplus 6A_1 \bigr] =  E_6 \bigoplus 6A_1,  \nonumber \\
&T_3 \circ \bigl[ E_6 \bigoplus 6A_1\bigr] = E_8\bigoplus \Pi^{3,1},
\label{action_of_Ts}
\end{align}
where $\circ$ refers to a composition operation and the Cartan matrices are given below. Also note that $6A_1 \equiv A_1 \bigoplus A_1 \bigoplus A_1 \bigoplus A_1 \bigoplus A_1 \bigoplus A_1$. 
\onecolumn
\begin{equation}
\biggl[ D_3 \bigoplus D_2\bigoplus A_1 \bigoplus 6A_1\biggr] =\begin{pmatrix}
2	&-1	&-1	&0	&0	&0	  \\
-1	&2	&0	&0	&0	&0	 \\
-1	&0	&2	&0	&0	&0	\\
0	&0	&0	&2	&-1	&0	 \\
0	&0	&0	&-1	&2	&0	 \\
0	&0	&0	&0	&0	&2	 \\
\end{pmatrix} \bigoplus \begin{pmatrix}
2	&0	&0	&0	&0	&0	  \\
0	&2	&0	&0	&0	&0	 \\
0	&0	&2	&0	&0	&0	\\
0	&0	&0	&2	&0	&0	 \\
0	&0	&0	&0	&2	&0	 \\
0	&0	&0	&0	&0	&2	 \\
\end{pmatrix} ,
\end{equation}
where $\begin{pmatrix}
a &b &0 &0\\
c & d &0 &0\\
0 &0 &p &q\\
0 &0 &m &n\\
\end{pmatrix} \equiv
\begin{pmatrix}
a &b \\
c & d
\end{pmatrix} \bigoplus \begin{pmatrix}
p &q \\
m & n
\end{pmatrix}$
\begin{equation}
\bigl[ E_4\bigr] = \bigl[A_4\bigr] = \begin{pmatrix}
2	&-1	&0	&0\\
-1	&2	&-1	&0\\
0	&-1	&2	&-1\\
0	&0	&-1	&2\\
\end{pmatrix},
\end{equation}
\begin{equation}
\bigl[ E_6 \bigoplus 6A_1 \bigr] = \begin{pmatrix}
2	&-1	&0	&0	&0	&0\\
-1	&2	&-1	&0	&0	&0\\
0	&-1	&2	&-1	&0	&-1\\
0	&0	&-1	&2	&-1	&0\\
0	&0	&0	&-1	&2	&0\\
0	&0	&-1	&0	&0	&2\\
\end{pmatrix}\bigoplus \begin{pmatrix}
2	&0	&0	&0	&0	&0	  \\
0	&2	&0	&0	&0	&0	 \\
0	&0	&2	&0	&0	&0	\\
0	&0	&0	&2	&0	&0	 \\
0	&0	&0	&0	&2	&0	 \\
0	&0	&0	&0	&0	&2	 \\
\end{pmatrix} ,
\end{equation}
\begin{equation}
\biggl[ E_8 \bigoplus \Pi^{3,1}\biggr]  = \begin{pmatrix}
2	&-1	&0	&0	&0	&0	&0	&0\\
-1	&2	&-1	&0	&0	&0	&0	&0\\
0	&-1	&2	&-1	&0	&0	&0	&0\\
0	&0	&-1	&2	&-1	&0	&0	&0\\
0	&0	&0	&-1	&2	&-1	&0	&-1\\
0	&0	&0	&0	&-1	&2	&-1	&0\\
0	&0	&0	&0	&0	&-1	&2	&0\\
0	&0	&0	&0	&-1	&0	&0	&2\\
\end{pmatrix} \bigoplus \begin{pmatrix}
2	&0	&0	&0 \\
0	&2	&0	&0 \\
0	&0	&2	&0\\
0	&0	&0	&2 \\
\end{pmatrix} ,
\end{equation}
The transformation matrices are listed as follows. 
\begin{equation}
T_1 = \begin{pmatrix}
    1        &0         &0         &0         &0         &0       &  0       &  0&0 &0\\
         0 &   1      &   0     &    0    &     0     &    0     &    0       &  0&0 &0\\
         0        & 0   & 1     &    0      &   0      &   0     &    0      &   0&0 &0\\
         0       &  0      &   0   & \frac{3}{2}     &    \frac{1}{4} &  \frac{3}{4}   &      0        & 0&0 &0\\
         0         &0        & 0      &   0  &  1     &  0    &     0      &   0&0 &0\\
         0         &0        & 0  & 1  & \frac{1}{2}   & \frac{3}{2}      & 0     &    0&0 &0\\
         0        & 0       &  0       &  0       &  0        & 0   & 1       &  0 &0 &0\\
         0        & 0        & 0        & 0       &  0        & 0     &    0   & 1	&0 &0\\
         0        & 0       &  0       &  0       &  0        & 0   & 0       &  0 &1 &0\\
         0        & 0        & 0        & 0       &  0        & 0     &    0   & 0 &0 &1\\
         \end{pmatrix} \bigoplus \begin{pmatrix}
         1 &0\\
         0 &1\\
         \end{pmatrix},
\end{equation}

\begin{equation}
\breve{T}_2 = \begin{pmatrix}
   \frac{3}{2}   & \frac{1}{4}   & \frac{3}{4}     &    0    &     0    &     0     &    0     &    0\\
   -\frac{1}{2} &   \frac{3}{4}  & -\frac{3}{4}   &  0    &     0  &       0    &     0       &  0\\
    \frac{1}{2}&  -\frac{1}{4}   & \frac{5}{4} & -\frac{2}{3}  & -\frac{1}{3}      &   0       &  0       &  0\\
   -\frac{1}{2}  & -\frac{1}{4}  & -\frac{3}{4}  &  \frac{4}{3}  &  \frac{2}{3}      &   0       &  0       &  0\\
         0   &      0       &  0    &     0         &0        & 0        & 0      &   0\\
         0    &     0         &0     &    0        & 0        & 0        & 0      &   0\\
         0    &     0       &  0       &  0       &  0      &   0       &  0      &   0\\
         0   &      0       &  0        & 0       &  0      &   0     &    0      &   0\\
\end{pmatrix} \bigoplus \begin{pmatrix}
         0 &0 &0 &0\\
         0 &0 &0 &0\\
         0 &0 &0 &0\\
         0 &0 &0 &0\\
         \end{pmatrix},
\end{equation}

\begin{equation}
T_2 = \begin{pmatrix}
\frac{3}{2}	&\frac{1}{4}	&\frac{3}{4}	&0	&0	&0	&0&	0\\
-\frac{1}{2}	&\frac{3}{4}	&-\frac{3}{4}	&0	&0&	0	&0	&0\\
\frac{1}{2}&	-\frac{1}{4}	&\frac{5}{4}	&-\frac{2}{3}	&-\frac{1}{3}	&-\frac{1}{2}	&0	&0\\
-\frac{1}{2}	&-\frac{1}{4}	&-\frac{3}{4}	&1	&0	&0	&0	&0\\
0	&0	&0	&0&	1	&0	&0	&0\\
-\frac{1}{2}	&-\frac{1}{4}&	-\frac{3}{4}	&0	&0	&1	&0	&0\\
0	&0	&0	&0	&0	&0	&1	&0\\
0	&0	&0	&0	&0	&0	&0	&1\\
\end{pmatrix}\bigoplus \begin{pmatrix}
         0 &0 &0 &0\\
         0 &0 &0 &0\\
         0 &0 &0 &0\\
         0 &0 &0 &0\\
         \end{pmatrix},
\end{equation}
and,
\begin{equation}
T_3 = \begin{pmatrix}
   1      &  0     &    0   &      0      &   0   &      0      &   0      &   0\\
   0 &   1    &     0       &  0       &  0  & 0   &      0      &   0\\
    1  &  2 &  4   & 2   & 1   & 2      &   0      &   0\\
    0   & 0   & 0 &  1 &   0  & 0        & 0        & 0\\
   -1 &  -2   &-3   &-2       &  0   &-2    &     0  & -\frac{1}{2}\\
    \frac{4}{3}    &\frac{8}{3}  &  4   & \frac{7}{3}   & \frac{2}{3}&    3 & -\frac{1}{2}    &     0\\
   -1  & -2  & -3   &-2  & -1  & -2  &  1     &    0\\
   -\frac{2}{3}  & -\frac{4}{3}  & -2   &-\frac{5}{3} &  -\frac{4}{3}  & -1       &  0 &   1\\
\end{pmatrix}\bigoplus \begin{pmatrix}
         1 &0 &0 &0\\
         0 &1 &0 &0\\
         0 &0 &1 &0\\
         0 &0 &0 &1\\
         \end{pmatrix}.
\end{equation}\\
\vspace{60 pt}
\begin{figure*}
\begin{center}
\includegraphics[scale=0.75]{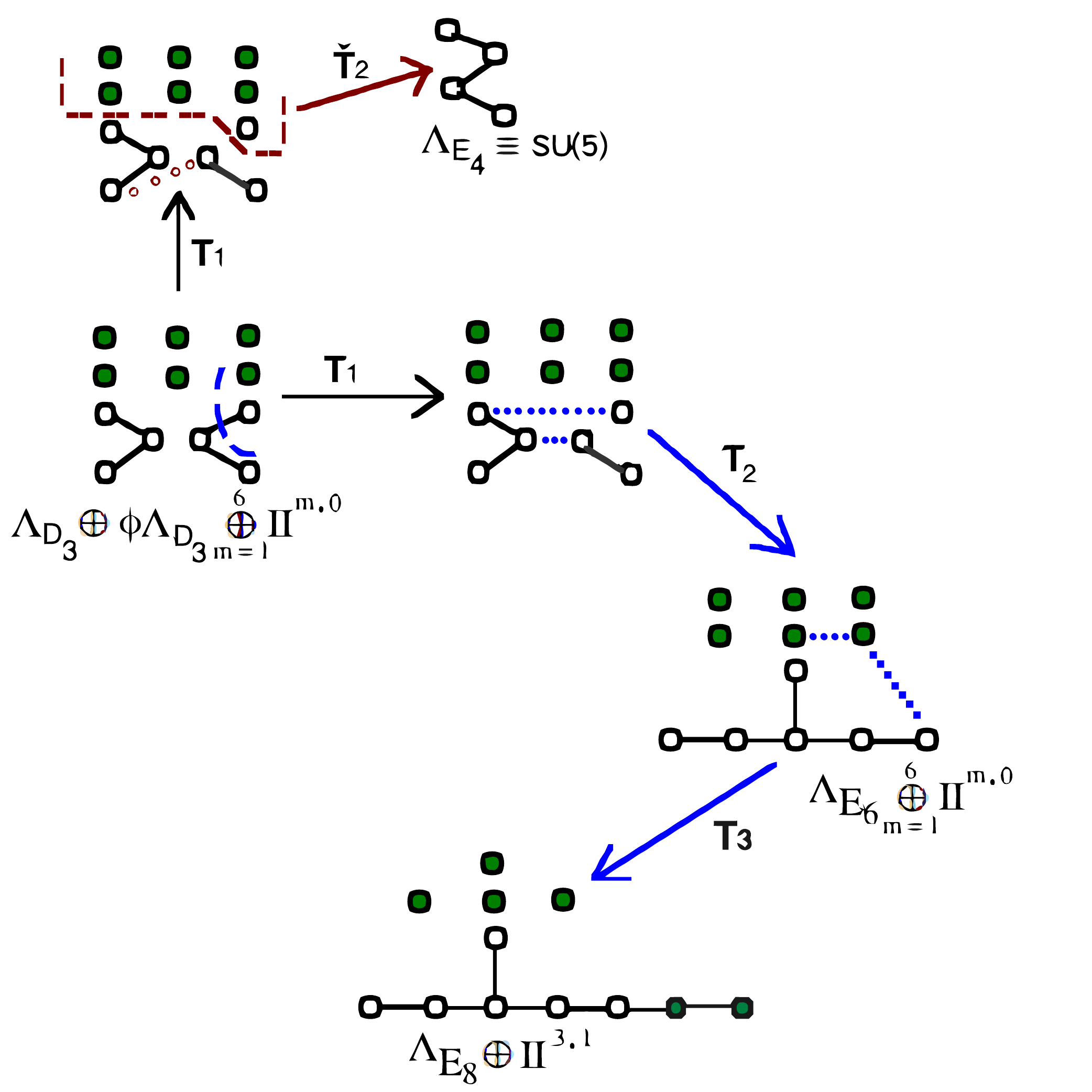}
\end{center}
\caption {A graphical representation of the emergence of lattices that constitute the unified worldspace beginning with the Dirichlet quantized host lattice, $\Lambda_{D_6^{3 \bigoplus \phi3}} \bigoplus \limits_{m=1}^6 II^{m,0} $. The solid dotted (dashed) lines denote the addition of edges between (removal of) nodes (edges and/or nodes). Note: In Dynkin graphs, nodes represent root vectors. Each hollow dotted line denotes the merging of two nodes. The solid green circles represent the six physical dimensions that are hidden in the internal space; subsequently, two of these hidden dimensions bridge the internal space and the physical Minkowski spacetime. The hollow circles represent nodes in the Euclidean (non-curved) space. The transformations $T_i, \breve{T}_2, ~i=\{1, 2, 3\}$ act on the corresponding Cartan matrices and regulate the emergence of higher dimensional lattices. }
\label{D6toE8}
\end{figure*}
\begin{multicols}{2}
The transformations given by the notation $T_i, ~i=1,2,3$ and $\breve{T}_2$ encode all the information about the evolution of the root-vectors and, hence, that of the emerging geometrical fabric of the spacetime coordinates. In other words, they regulate the setting of the Dirichlet quantized host $\Lambda_{D_3} \bigoplus \phi \Lambda_{D_3} $ thereby determining which geometry manifests itself. 
\subsection{Spectral norm of transformation matrices}
The spectral norms of the transformation matrices are given as follows,
\begin{align}
&||T_1||_{\sigma} = 1.6182, \nonumber \\
&||\breve{T}_2||_{\sigma}=2.6257, \nonumber \\ 
&||T_2||_{\sigma}=2.7215, \nonumber \\
&||T_3||_{\sigma} = 11.1872,
\label{Tnorms}
\end{align}
and the spectral norms of the lattices are as follows:
\begin{align}
&\biggl|\biggl|\bigl[\Lambda_{D_3} \bigoplus \phi \Lambda_{D_3}  \times A_1 \times A_1\bigr]\biggr|\biggr|_{\sigma}=3.4142, \nonumber \\
&\biggl|\biggl|\bigl[D_3 \bigoplus D_2 \times A_1 \bigoplus A_1 \times A_1\bigr]\biggr|\biggr|_{\sigma}=3.4142 \nonumber \\ 
&\biggl|\biggl|\bigl[E_4\bigr]\biggr|\biggr|_{\sigma}=3.6180, \nonumber \\
&\biggl|\biggl|\bigl[E_6\bigoplus A_1 \bigoplus A_1\bigr]\biggr|\biggr|_{\sigma} =3.9319, \nonumber \\ 
&\biggl|\biggl|\bigl[E_8\bigr]\biggr|\biggr|_{\sigma} = 3.989.
\label{Enorms}
\end{align}

\vspace{0.5 cm}

The norm of $T_3$ is an order of magnitude larger than the norms of all other transformation matrices. 
\subsubsection{Physical interpretation of the spectral norms}
Recall that the spectral norm of a transformation matrix reflects the largest singular value of the matrix that entails the maximum extent of allowable deformation (stretching) of vectors in the corresponding vector space \cite{Horn85}. Kleinert \cite{Kleinert89, Kleinert97, Kleinert00, Kleinert05} used a linear perturbation of the coordinate frame by a displacement field generating strain, and consequently revealed that space-time with torsion and curvature can be generated
from a flat Euclidean space-time using \textit{singular coordinate transformations}. He argued in favor of comparing the above to a crystallographic medium filled with dislocations and disclinations. Ruggiero and Tartaglia \cite{RT03} showed that Kleinert's singular coordinate transformations are the space-time equivalent of the plastic deformations which lead to incompatible defect states corresponding to generation of mass, mass current, and spin.

In our analysis presented above, $T_3$ is a representation in the Cartan sub algebra and coincides with the emergence of the 4-dimensional Minkowskii spacetime denoted by $II^{3,1}$. The aforementioned spike in the spectral norm of $T_3$ is a signature of the linear deformations\footnote{These singular value deformations are a combination of rotation and stretching/scaling processes.} in lattice geometry proposed by Kleinert to account for gravity. The spike in the spectral norm coincides with the link between the internal world, the Minkowskii physical world, and the curved spacetime denoted by the unimodular lattice $II^{3,1}$.

\section{Conclusion}
In this paper we have described a rigorous pathway for the emergence of 12D lattice from a 3D aperiodic substrate through an intermediary crystallographic host lattice of the internal space in 6D. The mathematical framework is set in the language of Clifford's geometric algebra and representations from Lie theory. We have demonstrated that the spawning of the new dimensions is inherently related to the symmetries of our base object viz. the 20G that has icosahedral symmetry. This has been possible by utilizing the concept of Dirichlet integers from the theory of algebraic rings. Consequently, we have presented a sequence of transformations to illustrate the emergence of spacetime geometry through the use of Dynkin graphs and Cartan algebra. The norms of these transformation matrices have a special significance in relation to the underlying geometry they induce.

Future work will include investigating the links between the quasicrystalline base and the higher dimensional geometry from a dynamical point of view. This will enable the possibility of mapping our understanding of dynamics in the measurable three dimensions with the dynamics encompassed by the $E_8$ Lie group. In a subsequent paper, we propose to perform detailed investigation of the Dirichlet host from the perspective of the theory of algebraic rings. This will shed further light on the importance of this intermediary host contributing toward a better understanding of the interconnections between group symmetries of higher dimensions.

\section{Acknowledgement}
The first author would like to acknowledge several useful suggestions and discussions with Dr. Carlos Castro Perelman. 

\end{multicols}
\section{Appendix: Vertex coordinates of the 20G, $(x+\phi x', y + \phi y', z+\phi z')$}
{\small $
\begin{array}{lll | l | lll | l}
\text{x} & \text{y} & \text{z} & \text{x+y+z} & \text{x$'$} & \text{y$'$} & \text{z$'$} & \text{x$'$+y$'$+z$'$} \\
\hline
0 & -2 & 2 & 0 & 0 & 0 & 0 & 0 \\

-2 & 0 & 2 & 0 & 0 & 0 & 0 & 0 \\

-2 & -2 & 0 & -4 & 0 & 0 & 0 & 0 \\

-2 & -1 & 1 & -2 & 1 & -1 & 0 & 0 \\

-1 & 1 & 0 & 0 & 1 & -2 & -1 & -2 \\

-1 & 0 & -1 & -2 & 2 & -1 & 1 & 2 \\

2 & -1 & -1 & 0 & -1 & -1 & 0 & -2 \\

1 & 1 & 0 & 2 & -1 & -2 & 1 & -2 \\

1 & 0 & 1 & 2 & -2 & -1 & -1 & -4 \\

2 & -1 & 1 & 2 & -1 & -1 & 0 & -2 \\

1 & -1 & 2 & 2 & 1 & 0 & -1 & 0 \\

1 & -2 & 1 & 0 & 0 & 1 & 1 & 2 \\

-1 & 1 & 0 & 0 & 1 & -2 & 1 & 0 \\

1 & 2 & 1 & 4 & 0 & -1 & 1 & 0 \\

0 & 1 & -1 & 0 & -1 & -1 & 2 & 0 \\

0 & 2 & -2 & 0 & 0 & 0 & 0 & 0 \\

-2 & 2 & 0 & 0 & 0 & 0 & 0 & 0 \\

-2 & 0 & -2 & -4 & 0 & 0 & 0 & 0 \\

2 & 1 & -1 & 2 & -1 & 1 & 0 & 0 \\

1 & 2 & -1 & 2 & 0 & -1 & -1 & -2 \\

1 & 1 & -2 & 0 & 1 & 0 & 1 & 2 \\

-2 & 1 & 1 & 0 & 1 & 1 & 0 & 2 \\

-1 & 2 & 1 & 2 & 0 & -1 & 1 & 0 \\

-1 & 1 & 2 & 2 & -1 & 0 & -1 & -2 \\

-2 & 1 & -1 & -2 & 1 & 1 & 0 & 2 \\

-1 & 0 & 1 & 0 & 2 & 1 & -1 & 2 \\

-1 & -1 & 0 & -2 & 1 & 2 & 1 & 4 \\

1 & -1 & 0 & 0 & -1 & 2 & -1 & 0 \\

2 & 1 & 1 & 4 & -1 & 1 & 0 & 0 \\
\end{array}$
$
\begin{array}{lll | l | lll | l}
\text{x} & \text{y} & \text{z} & \text{x+y+z} & \text{x$'$} & \text{y$'$} & \text{z$'$} & \text{x$'$+y$'$+z$'$} \\
\hline
1 & 0 & -1 & 0 & -2 & 1 & 1 & 0 \\

2 & -2 & 0 & 0 & 0 & 0 & 0 & 0 \\

2 & 0 & -2 & 0 & 0 & 0 & 0 & 0 \\

0 & -2 & -2 & -4 & 0 & 0 & 0 & 0 \\

-1 & -2 & 1 & -2 & 0 & 1 & 1 & 2 \\

1 & -1 & 0 & 0 & -1 & 2 & 1 & 2 \\

0 & -1 & -1 & -2 & 1 & 1 & 2 & 4 \\

1 & -2 & -1 & -2 & 0 & 1 & -1 & 0 \\

-1 & -1 & 0 & -2 & 1 & 2 & -1 & 2 \\

0 & -1 & 1 & 0 & -1 & 1 & -2 & -2 \\

1 & 0 & -1 & 0 & -2 & -1 & 1 & -2 \\

-1 & 1 & -2 & -2 & -1 & 0 & 1 & 0 \\

0 & -1 & -1 & -2 & -1 & 1 & 2 & 2 \\

-2 & -1 & -1 & -4 & 1 & -1 & 0 & 0 \\

-1 & -2 & -1 & -4 & 0 & 1 & -1 & 0 \\

-1 & -1 & -2 & -4 & -1 & 0 & 1 & 0 \\

2 & 0 & 2 & 4 & 0 & 0 & 0 & 0 \\

0 & 2 & 2 & 4 & 0 & 0 & 0 & 0 \\

2 & 2 & 0 & 4 & 0 & 0 & 0 & 0 \\

-1 & -1 & 2 & 0 & -1 & 0 & -1 & -2 \\

0 & 1 & 1 & 2 & -1 & -1 & -2 & -4 \\

1 & 0 & 1 & 2 & -2 & 1 & -1 & -2 \\

1 & -1 & -2 & -2 & 1 & 0 & 1 & 2 \\

0 & 1 & -1 & 0 & 1 & -1 & 2 & 2 \\

-1 & 0 & -1 & -2 & 2 & 1 & 1 & 4 \\

1 & 1 & 0 & 2 & -1 & -2 & -1 & -4 \\

0 & 1 & 1 & 2 & 1 & -1 & -2 & -2 \\

-1 & 2 & -1 & 0 & 0 & -1 & -1 & -2 \\

-1 & 0 & 1 & 0 & 2 & -1 & -1 & 0 \\

1 & 1 & 2 & 4 & 1 & 0 & -1 & 0 \\

0 & -1 & 1 & 0 & 1 & 1 & -2 & 0 \\
\end{array}
$}
\end{document}